\documentclass[article,11pt]{JHEP3}
\usepackage{epsfig}
\usepackage{amssymb}
\usepackage{graphics}
\usepackage{bbm}
\let\e=\epsilon         
      \let\l=\lambda  \let\m=\mu
\let\n=\nu                 
  \let\t=\tau      \let\f=\phi

     \let\L=\Lambda
           
\let\F=\Phi       \let\C=\Chi    
\def\brn{\bf}

\def\sp{\;\;,\;\;}

\newcommand{\Xc}{\cal X}
\newcommand{\Yc}{\cal Y}
\newcommand{\Zc}{\cal Z}
\newcommand{\Uc}{\cal U}
\newcommand{\Vc}{\cal V}
\newcommand{\Wc}{\cal W}
\newcommand{\Rc}{\cal R}
\newcommand{\Sc}{\cal S}
\newcommand{\Tc}{\cal T}
\newcommand{\be}{\begin{equation}}
\newcommand{\ee}{\end{equation}}
\newcommand{\bea}{\begin{eqnarray}}
\newcommand{\eea}{\end{eqnarray}}
\newcommand{\ba}{\begin{array}}
\newcommand{\ea}{\end{array}}
\def\nn{\nonumber}

\newcommand{\C}{{\mathbb C}}

\def\hri#1#2{\href{http://arxiv.org/abs/#1}{[ArXiv:#1]#2}}
\def\hre#1#2{\href{http://arxiv.org/abs/#1/#2}{[ArXiv:#1/#2]}}
\def\hspi#1#2{\href{http://www.slac.stanford.edu/spires/find/hep/www?irn=#1}{#2}}
\title{\Huge On mass hierarchies in orientifold vacua}
\author{\large Pascal Anastasopoulos$^{1,3}$
\footnote{Pascal.Anastasopoulos@roma2.infn.it},~~
\href{http://hep.physics.uoc.gr/~kiritsis/}{Elias Kiritsis}$^{2}$,~~
Andrea Lionetto$^1$\footnote{Andrea.Lionetto@roma2.infn.it},
\\ \\
$^1$ I.N.F.N.\ -\ Sezione di Roma ``Tor Vergata'', 00133, Roma, ITALY
\\ \\
$^2$ \href{http://hep.physics.uoc.gr/}{Department of Physics, University of Crete, 71003 Heraklion, GREECE}
\\ \\
$^3$ Department of Physics, CERN-Theory Division, 1211 Geneva 23, SWITZERLAND}
\preprint{0905.3044\\
ROM2F/2009/08\\
CERN-PH-TH/2009-063}

\abstract{We analyze the problem of the hierarchy of masses and mixings in orientifold realizations of the Standard Model.
We find bottom-up brane configurations that can generate such hierarchies. }
\keywords{D branes, Orientifolds, Standard Model, Mass hierarchy}

\begin{document}
\section{Introduction, motivation  and scope}
One of the biggest puzzles in the Standard Model (SM) is the origin and hierarchy
of masses and mixings.
When it comes to masses the scale of the problem is enormous: one needs to
explain a range of masses that spans fifteen orders of magnitude between the mass of the lightest
neutrino to the top. Moreover the patter of mixings is interesting. In the quark sector the first and second family
mix strongly while all other mixings are small. In the lepton sector, all mixings measured so far are maximal.
It seems to suggest that as we move up in mass mixings tend to become smaller.

There are several ideas on the origin of mass, the simplest being
via a Higgs scalar that is responsible for electroweak symmetry breaking.
This and related ideas are expected to be tested at LHC.
There are fewer and definitely less successful ideas that are purported
to explain the hierarchy of masses of the SM.
They can be roughly lumped into four classes: radiative mechanisms,
\cite{rad}, texture zeros, \cite{text,rr}, family symmetries \cite{family,fn,ir}
and seesaw mechanisms, \cite{seesaw}, although the classes are not
completely disjoint. In particular texture zeros can be considered as a class of family symmetries as
they are usually implemented via a discrete symmetry. Many of the ideas developed to deal with the mass hierarchy of th SM
are reviewed in \cite{rev-m}.

String theory has emerged as an arena for unifying interactions, in
 the last few decades. Finding the SM in a string theory vacuum
has proved a difficult task especially when it comes to match to the
 SM pattern of masses. So far none of  the early ideas on mass hierarchies
has been successfully implemented in a string vacuum, although string
 inspired use of anomalous U(1)'s in that direction was advocated \cite{irges}.
Recently, a simple implementation of the Froggatt-Nielsen idea was
advocated in the context of F-theory, \cite{vafa}.
There have been however partial hierarchies in the SM spectrum that
were successfully implemented like the top hierarchy and neutrino masses in the
heterotic string using higher order couplings, \cite{ar,far,atr,la},
or the third family and neutrino masses using large dimensions, \cite{akrt}.

Two perturbative landscapes of string theory vacua have monopolized
attention in the past two decades. The first to be analyzed was the
heterotic landscape
deemed interesting because of its large and appealing gauge symmetry
and the simplest structure of its perturbative expansion.
Although a large set of vacua was found, some of them phenomenologically
promising, several difficulties hampered the search for a SM-like vacuum,
most of all the fact that the string theory input in vacuum construction
(generalized geometry) is
quite disjoint with the output (spectra, gauge groups, low energy interactions).
At the same time, indications suggested that the heterotic string would
need be in strong coupling in order for some effects to be compatible
with data. At the same time,  type-I theory emerged as a strong-coupling dual
of the heterotic theory and SM-searches started to look in open string theory vacua.

Open string theory vacua, alias orientifolds, \cite{bs,horava, PS},  provided a fresh new perspective
 in the search for the SM, \cite{rev-1}-\cite{rev-4}. They allowed a bottom-up approach, \cite{bu1,bu2}
to building the SM, by utilizing the geometrized language offered
by D-branes supporting the SM interactions and particles.

The algorithm can be described as follows. One first constructs a
type II ground state, that involves
a closed CFT describing the compactification.
Then an appropriate orientifold projection is applied on the
closed string sector.
An open string sector is subsequently constructed by populating
 the allowed boundary states of the bulk CFT.
This part of the algorithm should be thought of as inserting D-branes
in the closed string vacuum in a
way compatible with the 2d-dynamics. In particular the D-brane configuration is such that it guarantees local
(as opposed to global) stability.
At this stage one can engineer the gauge group and spectrum with
rather milder constraints than those that are imposed at the end.
Therefore  a lot of the model building choices are decided early on.
Moreover, in this context, one first constructs the SM family of branes,
defined as the collection of boundary states that give rise to the chiral SM particles.

Finally, once the SM stack has been engineered to one's satisfaction,
 the stringy, tadpole cancellation constraints are imposed. 
This can be  done by adding in a modular fashion a ``hidden sector", ie.
 one or more brane stacks, that typically do not include light
 observable-hidden strings.
The procedure stops when tadpoles are eventually canceled.
This procedure has been algorithmized for a large set of RCFT building
blocks, and used to provide large lists of SM-like orientifold vacua, \cite{dhs,adks}.

In \cite{kst} a class of orientifold vacua were studied, constructed
from six copies of the second Gepner model (k=2).
The original motivation was to study quasi-realistic vacua using CFT
building blocks that are free CFTs.
A very interesting feature of the vacua described in \cite{kst}, was
that the 3 SM families do not originate from the same D-branes.
This has important consequences because of the generic presence of
anomalous U(1) symmetries in orientifold vacua.

Anomalous U(1) symmetries are ubiquitous in orientifolds. It has been
argued early on \cite{bu1,akrt},
that any SM orientifold realization must have at least one
and generically three anomalous U(1) symmetries, that make the most
characteristic signature of orientifold vacua.
Their phenomenological implications are diverse, \cite{ak}-\cite{anto}.

Their most important property, that impacts importantly on the dynamics of
the D-brane stack is that they provide numerous selection rules on the
effective couplings. In particular, they may be responsible for the absence
of the $\mu$-term, Yukawa couplings, baryon and lepton violating couplings etc.
However, as anomalous U(1)'s are effectively broken as gauge symmetries, the
selection rules they provide need qualification.
As the breaking of the gauge symmetry happens via the mixing with RR forms,
the global U(1) symmetry remains at this stage intact.
There are two types of realizations of anomalous U(1) symmetries as global
symmetries. If D-terms force charged fields to obtain vev's
then the global U(1) symmetry is broken. If on the other hand no vev's are
generated the anomalous U(1) global symmetry remain intact in perturbation theory.

However, the story must change beyond perturbation theory for two reasons.
The first is that we do not expect exact (compact) global symmetries to survive
in a gravitational theory. The second (in agreement with the first) is that there are always non-perturbative
effects that violate the associated global symmetry. The argument is simple.
A U(1) transformation involves a shift of RR field. The associated D-instanton effect
which is charged under the same RR field (the Stuckelberg axion) will violate by definition
the associated global U(1) symmetry. The effect is a D-instanton effect, whose
field theory limit sometimes may admit a gauge instanton interpretation,
\cite{bfkov}-\cite{stringinst}.
Therefore, couplings a priori forbidden by anomalous U(1)'s can have three
potential fates: (a) Be generated by a vev
if the U(1)' is broken by a charged scalar vev. (b) Be generated by an
instanton effect, if there is an
instanton with the requisite number of zero modes associated with a given
coupling. (c) Remain zero as no vev or instanton can generate it.

In view of the discussion above we may appreciate why, segregating SM
families on different D-branes may provide non-trivial selection rules of
Yukawa couplings, generating
eventually a hierarchy of masses.
Indeed, in the vacuum studied in \cite{kst}, for one of the quark family,
no Yukawa couplings were allowed by the anomalous U(1) symmetries.
Therefore, the Yukawa's for this family, if generated at all, they must be
generated by D-instantons and have therefore a natural exponential
suppression with respect to the other two families.\footnote{An exponential
 suppression of Yukawa's can also happen because of world-sheet instanton effects.
In the  particular  case of vacua constructed from intersecting $D_6$ branes
 this idea was explored in \cite{cim}.}

 A pertinent question at this stage is : are masses and mixings of the SM calculable
in terms of a more fundamental theory (in the same sense that the energy spectrum of hydrogen is calculable)
or are they ``environmental parameters" that happen to have these values although there are other SM-like vacua
where their values are different.
Most physicists believe in the first possibility and it is fair to say that in the absence of convincing evidence for the second
it is the most appealing one. However in the last few years there is evidence, in the context of string theory
that many aspects of SM-like ground-states are not unique, but there is a large landscape of vacua with varying properties.
We will not have anything to say on this issue that goes beyond our efforts in this paper.
We do not pretend either to provide mechanisms that uniquely predict masses and mixings, but we explore how
the associated hierarchies could be accommodated in orientifold vacua.

In this paper we will explore different effects that are prone to generate
interesting hierarchies between
fermion masses. Our scope is exploratory: there will be no concrete models of
masses and mixings neither predictions/postdictions for experiment.
The goal is to identify D-brane configurations that are promising when it comes
to generating the fermion hierarchy. This is the problem
we address in this paper. The next step will be to  construct such interesting
D-bane configurations.

There are several effects that can produce  hierarchically different Yukawa-like couplings.
\begin{itemize}

\item Tree-level cubic Yukawa couplings. This is the generic case when such
couplings are allowed. Their coefficient depends in general on several
ingredients. It is always proportional to the ten-dimensional dilaton but
also internal volumes, and other
backgrounds fields (internal magnetic fields, fluxes) enter. They may be
correlated with the associated gauge couplings if the fields participating
come from overlapping D-branes. They may also be free of volumes if the branes
intersect at points. Such variations are enough some times to explain the
 mass hierarchy
inside a family. An example of this was presented in \cite{akrt} in model B.
There the tree-level Yukawa's are such that once the top mass is fixed,
the bottom
and tau masses follow. It is important that such couplings are in the perturbative
regime for the picture to be consistent.
Another possibility that we study is that tree level couplings respect a discrete
symmetry (that may be a local symmetry of the D-brane configuration).
In such a case small variations of the closed string moduli may lead to an
appropriate hierarchy of Yukawa couplings.

\item Higher order couplings. These are couplings that appear beyond the cubic
level. They necessarily involve more fields than the SM fields.
These extra fields must obtain an expectation value in order for an effective
Yukawa coupling  to be generated.
Then such couplings  compared to the previous case carry an extra factor of
$\left({\langle\phi\rangle\over M_s}\right)^n$ with $n$ a positive integer.
Depending on the compactification the string scale may be replaced by a
compactification scale.
If  $\langle\phi\rangle\ll M_s$ this generates a hierarchy in the associated
Yukawa coupling. On the other hand the regime
$\langle\phi\rangle\gg M_s$ is non-perturbative.

\item D-Instanton-generated couplings. Such couplings violate the anomalous U(1)
symmetries. They are suppressed by exponential instanton factors
of the form $e^{-1/g}$ where $g$ is linearly related to the ten-dimensional
coupling constant and depends also on the volume of the cycle
the D-instanton is wrapped-on, as well as on other data (magnetic fields, fluxes etc).
In the particular case of gauge instantons $g$ is the square of the associated
gauge coupling.
In the well-controlled regime, $g\ll 1$ and multi-instantons are suppressed.
beyond the instanton-action factor, instanton-generated couplings carry a
characteristic scale. This is determined by the string scale, or other volume factors
affecting the world-volume factor of the D-instanton. Finally there is a one-loop
determinant that is generically of order ${\cal O}$(1).

\end{itemize}

In this paper we will explore structures that allow exploiting  a combination of
the couplings above to generate mass and mixing hierarchies.
One strategy will be the following:

\begin{enumerate}

\item We start from a D-brane configuration in the simplest bottom-up context, as first described in \cite{bu1} and generally defined in \cite{adks}.
It is described by  a set of SM and (anomalous) U(1) charges for the SM particles,
following the rules of D-brane engineering. In particular, generalized anomaly cancellation is imposed.
All cubic Yukawa couplings allowed by the gauge symmetries are considered non-zero.
We search and consider only bottom-up configurations that allow only one non-zero Yukawa coupling in each of the Up and Down quark
$3\times 3$ mass matrices.
The overall scale of masses is set by the vev's of the two electroweak Higgses $H_u$ and
$H_d$.\footnote{Orientifold realizations of the SM
even in the absence of supersymmetry, necessitate the presence of at least two Higgses $H_u$ and $H_d$
 with different charges under the Chan-Paton (CP) group. The reason is that the Higgs
 carries always an extra U(1) charge, associated typically to an anomalous U(1).
 The U and D quarks always have different values of such a U(1) charge in
 order to accommodate their difference in hypercharge. Therefore they couple to Higgses
 with different such U(1) charges. }

\item  Apart from the SM particles and Higgses, one more scalar $\Phi$ will be advocated to help with the generation of higher order Yukawa couplings.
Its vev will be selected to fit appropriate masses.

\item If a given Yukawa coupling is still zero, then a instanton contribution is advocated.
Different Yukawa couplings generated by the same instanton (same violation of U(1) charges) will be considered to have the same exponential factor.
This is stricter that what could really happen, as the same instanton can wrap two different cycles with very different volumes
and can thus generate very different exponential factors.
We will not use however this option in this paper.
We will choose the exponential factors at will to reproduce the masses.

\item The rest of the coefficients in the mass matrices are dimensionless couplings that we will assume to be of the same order
of magnitude and we will take them ad-hoc to vary in the interval [0.1,0.5].
A configuration will be deemed promising if it can reproduce the masses and mixings
of the SM with dimensionless couplings in that range.

\end{enumerate}

Such a strategy rests on a set of choices that could be otherwise.
For example sometimes couplings can be much smaller than the
range we choose. We do not pretend that our choices are universal.
They provide however a general first assessment of D-brane vacua
as to their ability to
generate multiple scales for masses and mixings.

It should be noted that the previous context for generating the mass
hierarchies of the SM, does not rest on family symmetries.
As was first analyzed in \cite{adks}, potential continuous family symmetries in
the context of orientifold vacua are very different
from those that have been explored in the QFT literature.
The reason is simple: the doublet-triplet of quarks is constrained to have its two end-points on the SU(3) and SU(2) stacks of branes.
Therefore the only extra charges it can carry are the $U(1)_3$ of the SU(3) charge (it is always present and it baryon number)
and potentially the $U(1)_2$ of the SU(2) in the case of a complex weak stack (this U(1) is not present if the group is Sp(2)).

In the latter case of real weak stack, there is absolutely no difference between
the three doublet-triplets, and they can carry no extra charges.
In the first case of a complex weak stack, the doublet-triplets can be distinguished
by the $U(1)_2$ charge that can take two possible values, $\pm 1$.
Again at most one doublet-triplet can be different from the other two. No non-abelian
charges are allowed in either case.

Even for discrete family symmetries the situation is different. In previous implementations such discrete symmetries come in two copies
acting on the
whole family  on the left and on the right (see for example  \cite{family}).
Here they typically come in one copy. A representative example are the discrete symmetries
that appear when branes are stuck at orbifold singularities
\cite{bu2}. We will explore the impact of such symmetries on the mass spectrum later on in this paper.

Our results are as follows: in section 3.1 we found all possible textures of the mass matrices for the quarks and leptons for brane configurations with three, four and five stacks of branes\footnote{It is worth mentioning that the five-brane stack configurations realize the most general mass form. Thus, we do not continue our analysis on vacua with six or more D-brane stacks.}. As stated above we aim to find among the possible orientifold vacua all models which give mass matrices with all the three scales.
We found no such solutions in the case of three brane stacks.

For four brane stacks we found (in section 4) a vacuum in which the highest mass scale is related to Yukawa terms, the intermediate mass scale to instantons while the lowest scale to higher order terms. The CKM matrix computed for this model agrees with the experimental result. We also found a vacuum which satisfies the CKM constraint but only with Yukawas and higher order terms in the lepton mass matrices. In this model, there is no 1-1 correspondence between the fermion masses in each family and the Yukawa, higher order and instantonic terms.
In the particular case of the KST model \cite{kst} we found a vacuum with only Yukawas and instantons which does not satisfy the CKM constraint.

Finally in the five stacks case we found a vacuum with three mass scales both in the quark and lepton sector which satisfies the CKM constraint.

The plan of our paper is as follows:
In section 2 we give the description of D-brane configurations that successfully realize
the SM spectrum.
In section 3 we study the general form of the mass matrices for the quarks and
leptons that is allowed in various configurations with three, four and five stacks of branes.
In section 4 we concentrate on vacua with four and five branes with mass matrices with
all the three scales.
In section 5 we concentrate on an orientifold vacuum with a $Z_3$ discrete symmetry and
we analyze the mass generation mechanism.
In section 6 we present our conclusions.

In the appendix we provide more details about the three, four, and five brane stack vacua.
We also provide the mass matrices for the quark, leptons and neutrinos of several
bottom-up configurations.

\section{Bottom-up description of D-brane configurations}

A D-brane realization of the SM requires several stacks of branes.
The minimum number of stacks is three \cite{bu1} and all three-stack realizations
were classified in \cite{AD}.
Most common realizations utilize four stacks. There are also realizations with a
higher number of stacks (an example is given in \cite{Honecker:2004kb}).

All such configurations have in common a unitary stack of three branes
(the ``color" or A stack), a stack of two branes (the ``weak" or B stack)
and then various numbers of extra branes. In the simplest case they can be
taken as single branes, but non-abelian stacks are also possible provided
the associated gauge symmetry is eventually  broken. Although this was explored in
\cite{adks} we will not entertain this possibility here. We will only consider two extra stacks,
C and D each made up of a single (complex) brane.

 Gauge fields are described by open strings with both endpoints on the same stack and,
  generically, they give
rise to Unitary, USp and SO groups. In particular the weak stack may have a U(2) or
Sp(2) group. We will assume a U(2) group, and we will mention at the end differences in the Sp(2) case.

The rest of the SM particles are open
strings attached on different (or the same) stack
providing bi-fundamental (as well as symmetric or antisymmetric) representations.
The hypercharge is a linear combination of the abelian factors of each stack.
Typically the other linear
combinations of the abelian factors are anomalous\footnote{$B-L$ in some cases may not be anomalous.}. These anomalies are canceled
by the Green-Schwarz mechanism and by generalized Chern-Simons terms\cite{gcs}.
The anomalous U(1) gauge bosons are massive and their masses can vary between the string scale or
much lower depending on appropriate volume factors \cite{u1mass}.

The quark doublets $Q$ are described by strings with one end on the `A-stack  and the other on the B-stack of branes.
The quark singlets $U^c, D^c$ are described either by strings which are stretched between stack A
and the two extra U(1)  C and D stacks.
It is also possible to be generated  by strings with both ends on the ``color"-brane.
In this case they transform in the antisymmetric representations of $SU(3)$ which is equivalent to the anti-fundamental.
The lepton doublets $L$ are described by strings which are stretched between the B stack and the C,D stacks while
the lepton singlets $E^c$ are described either by strings that are stretched between stacks C,D or by strings
with both ends on the same single brane (B,C,D). In this case they transform in a  symmetric representation of the corresponding abelian factor.
The right-handed neutrinos $N^c$ being SM singlets are either described by strings attached on the SM-branes or they may come from the hidden
sector of the model. In the first case they can be either stretched between the C,D stacks or they might have both ends
on the B brane. In such a  case they transform under the antisymmetric representation of the $SU(2)$ which is equivalent to the singlet.

\section{Mass Matrices of the SM stack}

Our main interest is to study the mass generation mechanism in orientifolds. As we mentioned above the
SM particles\footnote{Our statements in this section do not assume spacetime  supersymmetry.}
are described by open strings whose ends are attached on various stacks of branes.
In this case many Yukawa terms are forbidden due to the fact that they are not gauge invariant
under the appropriate U(1) symmetries.

\FIGURE[t]{\epsfig{file=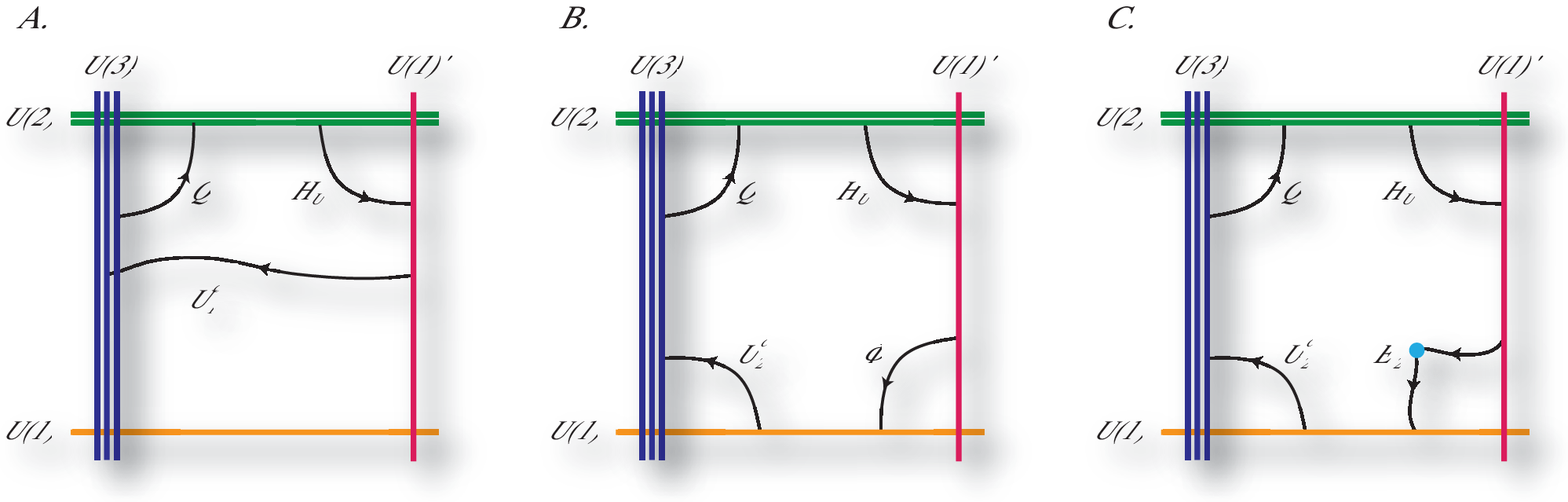,width=15cm}\caption{The three types of
mass generating terms: The configuration A allows for a Yukawa term. However, in the B and C cases
no Yukawa terms can be generated. In the B case there is a higher order term due to
the presence of a field $\F$, while in the C case there is a contribution from an instanton term $E_2$. \label{Picture}}}

An example of the four stacks' case is sketched in figure \ref{Picture}, where $Q\sim (1,-1,0,0)$,
$U^c_1\sim (-1,0,0,1)$,
$U^c_2\sim (-1,0,1,0)$ and $H_u\sim (0,1,0,-1)$\footnote{The notation $(q_A,q_B,q_C,q_D)$ indicates the U(1) charges of a state under the four
diagonal U(1) symmetries of the four D-brane stacks. The A stack contains the color U(3) group. The B stack contains the weak U(2) group.
The B and C stacks are assumed to have U(1) groups.}. The Yukawa term $Q U^c_1 H_u$ is uncharged
under the four abelian factors and therefore is allowed while the term $Q U^c_2 H_u$ has charge
 $(0,0,1,-1)$ and thus forbidden. Such term could not contribute to the quark mass matrix.
Here  we will entertain the possibility that there are  non-zero contributions for these mass matrix
entries from  higher-order terms and non-perturbative contributions.

Higher order terms must contain fields which are not present in the SM spectrum (with the exception of the neutrino mass terms).
It is a generic feature of stringy spectra D-brane that  additional non-chiral fields
are present. In orientifolds some of them are important for generalized anomaly and tadpole cancellation.
These fields can provide higher-order terms in the mass matrices.
For example, we consider one of these additional field $\f_1$ with charges $(0,0,-1,1)$, a SM singlet, originating
from the non-chiral part of the spectrum. We also assume that $\f_1$ acquires a non-zero vev $V_{\f_1}$.
In this case, the effective action contains a higher-order  term of the form
$Q U^c_2 H_u\f_1$ that provides a quark mass term proportional to $V_u V_{\f_1}/M_s$
where $V_u$ is the vev of  $H_u$ and $M_s$ is the string scale.
Since $V_{\phi_1} /M_s \lesssim1$ in the perturbative regime, such a contribution is smaller
than the leading Yukawa term.
For the same reason, higher-order terms that are in principle allowed are suppressed by higher powers of $M_s$.
Such scale differences in the mass matrices could be used to explain  the hierarchy between the fermion masses.\footnote{And indeed it was used in
\cite{ar,far,atr}.}

When neither Yukawa nor higher order terms are allowed, we may consider non-perturbative contributions.
D-instanton contributions give Yukawa couplings of the form  $Q U^c_2 H_u ~e^{- S_I}$
where $I$ denotes the type of instanton and the action, the coefficient $e^{- S_I}$ indicates the instanton action,
$S_I$ is proportional to  the internal volume the instanton brane wraps and it may also depend on other closed string moduli,
\cite{bfkov,i-review,stringinst}.
For internal volumes a few times the string scale such contributions are exponentially suppressed.
Summarizing, the following Yukawa-like terms can contribute to the mass matrix:
\begin{itemize}
\item Yukawa terms of the form $g_i Q u H_u$
\item Higher order terms of the form $g_i QuH_u \f /M_s$ where $\f$ a scalar field with
zero hypercharge. Such terms are suppressed by the string scale $M_s$.

\item Instanton terms of the form $g_i~ Q u H_u \times e^{- S_I}$.
We will assume that  $e^{- S_I}\lesssim1$ so that we can neglect multi-instanton terms.

\end{itemize}
In all the previous terms the $g_i$'s are dimensionless coupling constants, which we
assume to be of the same order ${\cal O}(1)$ and in the perturbative regime\footnote{In  practice and for concreteness we will assume them to  take values
between 0.1 - 0.6 although the precise bounds are also a matter of taste.}.

\subsection{Mass Matrix Forms}

In this section we study the general form of the mass matrices for the quarks and leptons that is allowed in various configurations with three, four and five stacks of branes.

We consider only vacua with two Higgs
doublets  $H_u,~ H_d$ which in particular could also accommodate the MSSM.

As mentioned in the introduction, in all orientifold  vacua  either all quark doublets are described by the
same type of charges ($Q_1=Q_2=Q_3$) or one quark doublet is different from the other
two $Q_1\neq Q_2=Q_3$. Therefore, as far as U(1) selection rules are concerned either all rows in the mass matrix
will have the same type
or one will be different from the other two.

After studying all possible bottom-up brane configurations we find that the resulting quark mass matrices are of the following form:
\footnote{Some of the entries below may be zero, compatible with associated formats. This is an interesting possibility which we will not however pursue in this paper.
We will only note the mass-matrix zeros cannot be of the type discussed in the earlier literature as they have to be compatible with the forms below.}
\bea
M_{Form-1} &=&~\left(
\begin{array}{lll}
\Xc ~ &\Xc ~ &\Xc  \\
\Xc ~ &\Xc ~ &\Xc  \\
\Xc ~ &\Xc ~ &\Xc  \\
\end{array}\right)\label{form1}\\
M_{Form-2} &=&~\left(
\begin{array}{l | ll}
\Xc ~ &\Yc ~ &\Yc  \\
\Xc ~ &\Yc ~ &\Yc  \\
\Xc ~ &\Yc ~ &\Yc  \\
\end{array}\right)\label{form2}
~\sim
~\left(
\begin{array}{lll}
\Xc ~ &\Xc ~ &\Xc  \\
\hline
\Yc ~ &\Yc ~ &\Yc  \\
\Yc ~ &\Yc ~ &\Yc  \\
\end{array}\right)\\
M_{Form-3} &=&~\left(
\begin{array}{l | ll}
\Xc ~ &\Yc ~ &\Yc  \\
\hline
\Zc ~ &\Uc ~ &\Uc  \\
\Zc ~ &\Uc ~ &\Uc  \\
\end{array}\right)\label{form3}\\
M_{Form-4} &=&~\left(
\begin{array}{l | l | l}
\Xc ~ &\Yc ~ &\Zc  \\
\Xc ~ &\Yc ~ &\Zc  \\
\Xc ~ &\Yc ~ &\Zc  \\
\end{array}\right)\label{form4}\\
M_{Form-5} &=&~\left(
\begin{array}{l | l | l}
\Xc ~ &\Yc ~ &\Zc  \\
\hline
\Uc ~ &\Vc ~ &\Wc  \\
\Uc ~ &\Vc ~ &\Wc  \\
\end{array}\right)\label{form5}
\eea
where $\Xc,\Yc,\Zc,\Uc,\Vc,\Wc$ denotes terms of the same type, either Yukawa,
higher-dimension  or instantonic terms.
While there can be only one kind of Yukawa and higher-dimension  terms,
in general there can be several different instantonic
terms\footnote{This arises because there could be several instantons contributing, wrapping different compact
internal cycles and therefore giving contributions of different size.}.
For example there are vacua in which the $\Zc$'s and the $\Uc$'s in~(\ref{form3})
are all instantons but they are different from each other. Specific examples will be given in the
following sections.
Note that we consider as equivalent the two matrices (\ref{form2}) since they have the same hierarchy in  their
eigenvalues.

In the lepton sector, in addition to (\ref{form1}-\ref{form5}) we can also have vacua
where all the entries in the mass matrix are different:
\bea
M_{Form-6} &=&~\left(
\begin{array}{l | l | l}
\Xc ~ &\Yc ~ &\Zc  \\
\hline
\Uc ~ &\Vc ~ &\Wc  \\
\hline
\Rc ~ &\Sc ~ &\Tc  \\
\end{array}\right)\label{form6}\eea
Note that there are vacua in which the ``weak" stack provides an $Sp(2)$ instead
of $U(2)$. Since $Sp(2)$ is isomorphic to $SU(2)$ we do not have in
this case the $U(1)$ factor associated to this stack of branes.
Therefore, the quark doublets which are stretched between the
A and the B branes have the same charges (+1,0,0) under the $U(1)_A\times U(1)_C\times U(1)_D$ and therefore the three doublets
have all the same U(1) charges.
In this case, the quark mass matrices can only have one of the forms: (\ref{form1}, \ref{form2}, \ref{form4}).


Below we give a classification of the D-brane realizations of the SM considering
the possible form of the quark mass matrices.
We restrict this analysis to the quark sector since in the lepton sector all mass matrix forms
(\ref{form1}-\ref{form6}) are allowed in each realization.

\subsection*{Three Stacks: the $U(3)\times U(2)\times U(1)$ realizations}

This setup has been first considered in detail in~\cite{AD}.
In this case there are two classes of vacua characterized by two different hypercharge embedding:
$Y=-{1\over 3}Q_{\brn a}-{1\over 2}Q_{\brn b}$ and $Y={1\over 6}Q_{\brn a}+{1\over 2}Q_{\brn c}$
(the charge assignments for all the SM fields is given in appendix \ref{3branes}).
\begin{itemize}

\item For $Y=-{1\over 3}Q_{\brn a}-{1\over 2}Q_{\brn b}$, the only possible form for both quark mass matrices
$M_U$ and $M_D$ is (\ref{form1}).

\item For $Y={1\over 6}Q_{\brn a}+{1\over 2}Q_{\brn c}$, there are two different possible charge assignments
for the $d$-quarks allowing the corresponding mass matrix to be of the form either (\ref{form1}) or (\ref{form2}).
\end{itemize}

\subsection*{Four Stacks: the $U(3)_A\times U(2)_B\times U(1)_C\times U(1)_D$ realizations}

In this case, there are seven different hypercharge embeddings (see appendix \ref{4branes}).
\begin{itemize}

\item For $Y=-{1\over 3}Q_{\brn a}-{1\over 2}Q_{\brn b}+Q_{\brn d}$, both $M_U,~M_D$ can be of the form
(\ref{form1}, \ref{form2}). This hypercharge embedding was identified as  model A in~\cite{akrt}.

\item For $Y={2\over 3}Q_{\brn a}+{1\over 2}Q_{\brn b}+Q_{\brn c}$, $M_U$ can be of the form
(\ref{form1}, \ref{form2}) while $M_D$ can only be of the form (\ref{form1}).
This hypercharge embedding was identified  as model B in~\cite{akrt}.

\item For $Y={1\over 6}Q_{\brn a}+{1\over 2}Q_{\brn c}-{1\over 2}Q_{\brn d}$ (also known as  the Madrid embedding, \cite{madrid}),
$M_U$ can be of the form (\ref{form1}-\ref{form3}) while $M_D$ can be of the form (\ref{form1}-\ref{form5}).

\item For $Y={1\over 6}Q_{\brn a}+{1\over 2}Q_{\brn c}-{3\over 2}Q_{\brn d}$, $M_U$ can be of the form
(\ref{form1}, \ref{form2}) while $M_D$ can be of the form (\ref{form1}-\ref{form3}).

\item For $Y=-{1\over 3}Q_{\brn a}-{1\over 2}Q_{\brn b}$, $M_U$ can only be of the form
(\ref{form1}) while $M_D$ can be of the form (\ref{form1}, \ref{form2}, \ref{form4}).

\item For the last two embeddings\footnote{These embeddings were never found in the extensive search of \cite{adks}.
Their existence is therefore in doubt.} $Y=-{5\over 6}Q_{\brn a}-{Q_{\brn b}}-{1\over 2}Q_{\brn c}+{3\over
2}Q_{\brn d}$ and
$Y={7\over 6}Q_{\brn a}+{Q_{\brn b}}+{3\over 2}Q_{\brn c}+{1\over
2}Q_{\brn d}$ both $M_U,~M_D$ are of the form (\ref{form1}).

\end{itemize}
It is worth noting that in the Madrid embedding we get the highest
number of different types of  mass matrices. This is due to the fact that the C and D
stacks contribute equally to the hypercharge allowing many alternative configurations.
In the Madrid class of vacua it is possible to have charge assignments such that for the quarks, $Q_1\neq Q_2=Q_3$,
$U_1\neq U_2=U_3$ and $D_1\neq D_2\neq D_3$. This in turn implies that
both $M_U$ and $M_D$ can have three different kind of entries.
In all the other vacua we considered, at least one of the following charge assignment,
either $Q_1=Q_2=Q_3$ or $U_1=U_2=U_3$ or $D_1= D_2= D_3$, is realized.

\subsection*{Five Stacks: the $U(3)_A\times U(2)_B\times U(1)_C\times U(1)_D \times U(1)_E$ realizations}

In this framework, we found 23 possible hypercharge embeddings. Among these embeddings, 12 of them
have either $M_U$ or $M_D$ or both on them of the form (\ref{form1}).
8 of them
have either $M_U$ or $M_D$ or both on them of the form (\ref{form1}) or (\ref{form2}).
The remaining three are the most interesting ones where the mass matrices $M_U$ and $M_D$ can have at least
three scales:
\begin{itemize}
\item For $Y={1\over 6}Q_{\brn a}+{1\over 2}Q_{\brn c}-{1\over 2}Q_{\brn d}-{3\over 2}Q_{\brn e}$ and
$Y={1\over 6}Q_{\brn a}+{1\over 2}Q_{\brn c}-{1\over 2}Q_{\brn d}$, $M_U$ can be of the
form (\ref{form1}-\ref{form3}) while $M_D$ can be of the form (\ref{form1}-\ref{form5}).
\item For the ``Madrid-like" 5 stacks extension: $Y={1\over 6}Q_{\brn a}+{1\over 2}Q_{\brn c}
+{1\over 2}Q_{\brn d}+{1\over 2}Q_{\brn e}$ where all single branes democratically contribute with
a factor 1/2 to the hypercharge, both $M_U$ and $M_D$ can be of the form (\ref{form1}-\ref{form5}).
\end{itemize}
Note that as in the four-stacks' case,  the Madrid-like embedding gives the highest number of different mass matrices.
In addition, only in this context we can have both $M_U,~M_D$ of the form (\ref{form5}).

\section{Vacua with 3-scales in all fermion mass matrices}

In each of the SM mass matrices for the quarks and leptons there is a large hierarchy.
In this work, we want to explore  the possibility that the different scales are related to
the three different types of possible Yukawa-like terms.
Therefore, we consider  vacua where the quark and lepton mass matrices have
the forms (\ref{form3}-\ref{form6}). This excludes all three stack constructions (\ref{3branes})
as well as all four stack constructions (\ref{4branes}) apart from a subclass of the
``Madrid" vacua (\ref{Madrid})\footnote{In \cite{leontaris, Cvetic:2009yh}, several other
 embeddings of the MSSM in D-brane configurations have been analyzed with focus on Yukawa couplings, and masses.}.

In order to satisfy the above requirements, the vacuum should contain one quark doublet different from the other two
(say $Q_1\neq Q_2=Q_3$) as well as a different right-handed $U^c$ and $D^c$ from the other two
(say $U^c_1\neq U^c_2=U^c_3$ and $D^c_1\neq D^c_2=D^c_3$). This choice ensures the ``3-scales" in the quark
mass matrices and fixes all the quarks since each of them has only two possible descriptions.
For the choice of leptons we have some more freedom since each lepton doublet and singlet can get several different
charge assignments (as can be seen in (\ref{Madrid})).
A subclass of the Madrid vacua that satisfy our requirements is:
\bea
&&Q_1~~~~: ~(~~1, + 1, ~~0, ~~0) ~~~,~~ Q_2,Q_3~:~(~~1, - 1 ,~~ 0 ,~~ 0)\nn\\
&&U^c_1~~~~:~(-1, ~~0 , -1, ~~0) ~~~,~~ U^c_2~U^c_3~:~(-1,~~ 0 , ~~0 , -1)\nn\\
&&D^c_1~~~~:~(-1, ~~0 , +1, ~~0)~~~,~~ D^c_2~D^c_3:~(-1, ~~0 ,~~ 0 , +1)\nn\\
&&L^c_1~~~~:~(~~0, ~~1 , -a , a-1) ~,~ L^c_2~L^c_3~:~(~~0, ~~1 , -b, b-1)\nn\\
&&E^c_1~~~~:~(~~0, ~~0 , ~~c , ~~d) ~~~,~~ E^c_2~~~~~:~(~~0, ~~0 , ~~e, ~~f) ~~,~~ E^c_2~:~(~~0, ~~0 ,~~ g, ~~h)\nn\\
&&N^c_{1,2,3}:~(~~0, \pm 2 , ~~0, ~~0)
\label{vacua1}
\eea
where $a,b=(0,1)$ and $c,d,e,f,g,h=(0,1,2)$ with the constraint $|c+d|=|e+f|=|g+h|=2$.
Notice that the lepton doublets can all have different charge assignments in a single vacuum.
The two MSSM Higgses are described by
\bea
H_u~&:&~ (0, -1 , +1 , 0)\nn\\
H_d~&:&~ (0, -1 , -1 , 0)
\eea
and we consider two additional scalars with zero hypercharge $\f_1$ and $\f_2$, coming from the non chiral part of the spectrum:
\bea
\f_1~&:&~ (0, 0 , -1 , +1)\nn\\
\f_2~&:&~ (0, 0 , +1 , -1) ~.
\eea

The brane configurations that we consider here are subject to two constraints: the
spectrum must match that of the MSSM in the chiral sense, with chirality
defined with respect to $SU(3) \times SU(2) \times U(1)$. Furthermore all cubic anomalies in
each factor of the full Chan-Paton group must cancel. This must be true because we
want to be able to cancel tadpoles, and tadpole cancelation imposes cubic anomaly
cancelation (mixed anomalies are canceled by the generalized Green-Schwarz mechanism).
The tadpoles are usually canceled by adding hidden sectors, which adds
new massless states to the spectrum. These are non-chiral and thus they do not
alter the anomaly cancelation mechanism in the MSSM sector.
As described in \cite{adks} the
cubic anomaly cancelation conditions that are derived from tadpole cancelation are
the usual ones for the non-abelian subgroups of $U(N)$, $N > 2$. Vectors contribute 1,
symmetric tensors $N+4$ and anti-symmetric tensors $N-4$, and conjugates contribute
with opposite signs. But the same condition emerges even if $N = 1$ and $N = 2$. This
means that for example a combination of three vectors and an anti-symmetric tensor
is allowed in a $U(1)$ factor. This is counter-intuitive, because the anti-symmetric
tensor does not even contribute massless states, so that one is left with just three
chiral massless particles, all with charge 1. The origin of the paradox is that it is
incorrect to call this condition ``anomaly cancelation" if $N = 1$ and $N = 2$ and if
chiral tensors are present. It is simply a consequence of tadpole cancelation; the
anomaly introduced by the three charge 1 particles is factorizable, and canceled by
the Green-Schwarz mechanism.

Using the above constraints we finally get
eight vacua which are anomaly free and with 3 different kind of terms in the quark and lepton mass matrices.
Inserting the values of table~\ref{8vacua} in (\ref{vacua1}) we get the different charge assignments for each vacuum.

In the appendix F we provide a complete description of the mass matrices.

\begin{table}[t]
\centering
\begin{tabular}[htb]{|c|c|c|c|c|c|c|c|c|}
\hline vacuum & $ ~~a ~~$ & $ ~~b ~~$  & $ ~~c ~~$ & $ ~~d ~~$ & $ ~~e ~~$& $ ~~f ~~$& $ ~~g ~~$& $ ~~h ~~$\\
\hline
\hline 1:  & 1 & 1 & 0 & 2 & 1 & 1 & 2 & 0 \\
\hline 2:  &  0 & 1 & 0 & 2 & 2 & 0 & 2 & 0 \\
\hline 3:  &  1 & 0 & 0 & 2 & 1 & 1 & 1 & 1 \\
\hline 4:  &  1 & 0 & 1 & 1 & 2 & 0 & 2 & 0 \\
\hline 5:  &  0 & 1 & 0 & 2 & 0 & 2 & 1 & 1 \\
\hline 6:  &  0 & 1 & 1 & 1 & 1 & 1 & 2 & 0 \\
\hline 7:  &  1 & 0 & 0 & 2 & 0 & 2 & 2 & 0 \\
\hline 8:  &  0 & 0 & 0 & 2 & 1 & 1 & 2 & 0 \\
\hline
\end{tabular}
\caption{The eight consistent vacua with 3-scales in each of the mass matrices of the quarks and leptons.}\label{8vacua}
\end{table}

\subsection{Vacuum 1: $a=1, ~b=c=e=h=0, ~d=f=g=2$.}

As an example, we choose the first vacuum in table \ref{8vacua}. The corresponding mass matrices for the quarks have the form:
\bea
M_U &=&V_u~\left(
\begin{array}{lll}
g_1 ~        &g_2 v_{\f_1} &g_3 v_{\f_1} \\
g_4 E_1 &g_5 E_2        &g_6 E_2 \\
g_7 E_1 &g_8 E_2        &g_9 E_2 \\
\end{array}\right)\label{Uquarkmass1}\\
M_D &=&V_d~\left(
\begin{array}{lll}
q_1 ~        &q_2 v_{\f_2} &q_3 v_{\f_2} \\
q_4 E_1 &q_5 E_3       &q_6 E_3 \\
q_7 E_1 &q_8 E_3       &q_9E_3 \\
\end{array}\right)
\eea
where $v_{\f_1}=V_{\f_1}/M_s$, ${v_\f}_2=V_{\f_2}/M_s$ and $E_i=e^{-Vol_{I_i} I_i}$ are the dimensionless instantons.
These two matrices have the form (\ref{form3}) where $\Xc$ is a Yukawa term, $\Yc$'s are higher terms and the $\Zc$'s and the $\Uc$'s are
instantons $E_1$ and $E_2$ respectively.
The above matrices are the same for all the eight vacua in (\ref{vacua1}).
The mass matrices for the leptons and neutrinos change form and in this specific vacuum we have:
\bea
M_L &=&V_d~\left(
\begin{array}{lll}
l_1 E_4 &l_2 v_{\f_1}  &l_3 ~ \\
l_4 E_4 &l_5 v_{\f_1}  &l_6 ~ \\
l_7 E_4 &l_8 v_{\f_1}  &l_9 ~ \\
\end{array}\right)\label{Leptonmass}\\
M_N &=&
\left(
\begin{array}{llllll}
0 & 0 & 0 & g_{11} V_u E_1 & g_{12} V_u E_1 & g_{13} V_u E_1 \\
0 & 0 & 0 & g_{21} V_u E_1 & g_{22} V_u E_1 & g_{23} V_u E_1 \\
0 & 0 & 0 & g_{31} V_u E_1 & g_{32} V_u E_1 & g_{33} V_u E_1 \\
g_{11} V_u E_1 & g_{21} V_u E_1 & g_{31} V_u E_1 & q_{11}M_sE_5 & q_{12}M_sE_5 & q_{13}M_sE_5 \\
g_{12} V_u E_1 & g_{22} V_u E_1 & g_{32} V_u E_1  & q_{21}M_sE_5 & q_{22}M_sE_5 & q_{23}M_sE_5 \\
g_{13} V_u E_1 & g_{23} V_u E_1 & g_{33} V_u E_1  & q_{31}M_sE_5 & q_{32}M_sE_5 & q_{33}M_sE_5
\end{array}\right)\label{Nmatrix}
\eea
where $g_i,~q_i,~l_i$ and $g_{ij},~q_{ij}$ are dimensionless couplings assumed
to be of the same order.\footnote{The tiny neutrino masses are generated through
 the seesaw mechanism. Schematically the terms that can contribute to the neutrino mass matrix have the following form:
\bea
g_{ij}L_i N_j^c H_u + q_{ij} M_s N^c_i N^c_j
\eea}
It is easy to check that neither Yukawa nor Majorana terms are present in the neutrino mass matrix
for the vacua of table~\ref{8vacua}. The only way to get such terms is by instantonic contributions
$E_1$ and $E_5$. The $E_1$ are the same dimensionless instantonic contribution that also
appear in the $U$-quark mass matrix (\ref{Uquarkmass1}).

The parameters of our vacuum are evaluated by equating the eigenvalues of all the mass
matrices to the running values of the quark, lepton and neutrino masses at various scales.
If the vacuum is supersymmetric, the low-energy effective action has softly broken Susy
and therefore all couplings run logarithmically. In the non-supersymmetric case, some
other solution to the hierarchy problem must be invoked so that couplings run logarithmically.

The values of the quarks, leptons and neutrino masses at various scales have been computed in the MSSM framework \cite{Xing:2007fb}.
Since there are several unknown parameters in each mass matrix, we fix some of them and we solve the system for the remaining ones.
We perform this task by requiring that all the dimensionless couplings must be of the same order.

For example, in the $M_U$ matrix, all entries in the $2\times 2$ matrix:
\bea
\left(\begin{array}{ll}
g_5 ~E_2        &g_6 ~E_2 \\
g_8 ~E_2        &g_9 ~E_2 \\
\end{array}\right)
\eea
should be of the same order due to our constraint. This requirement is not in general satisfied in the MSSM.
In our analysis we explore three possibilities for the value of the string scale: $M_s=1$ TeV,
$M_s=10^{12}$ GeV and at $M_s=\L_{GUT}=2\times 10^{16}$ GeV scale.

Here we give a brief description of the strategy we followed in order to determine the values
of the vev's and instantons. Each mass matrix entry is parametrized as a product of a
coupling $g_i$ with one of the relevant parameter, namely a plain Yukawa term, a vev or
an instanton. We fix some of these parameters to reproduce for example the masses of the
heaviest quarks. The remaining parameters are fixed by imposing the equality of the mass
matrices eigenvalues with the experimental values. The couplings $g_i$ are used as fine
tuning parameters varying their values at random in a small interval $[0.1,0.6]$.

For the present vacuum (as well as for the second and third in table \ref{8vacua}) we were able to find solutions where
\bea
&&V_u\sim m_t,~V_d\sim m_b\nn\\
&&E_1\sim E_2 \sim m_c/m_t 
\nn\\
&&E_3\sim E_4\sim m_s/m_b 
\nn\\
&& v_{\phi_1} \sim m_u/m_t 
\nn\\
&&v_{\phi_2} \sim m_d/m_b 
\label{YIH}\eea
where $m_i$ are the masses of the corresponding quarks,
and all couplings $|g_i|$, $|q_i|$, $|l_i|$, $|g_{ij}|$, $|q_{ij}|$ are within the range $[0.1,0.6]$.
%
%

As we mentioned before, in order to get the tiny neutrino masses we have to implement the seesaw mechanism.
The main idea of this mechanism can be sketched in a simple case of a $2\times 2$ matrix:
\bea
M_N &\sim& \left(
\begin{array}{ll}
0 & m \\
m & M
\end{array}\right)\label{NmatrixGeneral}
\eea
where $M\gg m$.
This matrix has one eigenvalue which is proportional to $M$ while the other one is proportional to $m^2/ M$.
The previous result can be easily extended in our case of the $6\times 6$ neutrino mass matrix~(\ref{Nmatrix}).
In this matrix, all the entries of the off diagonal $3\times 3$ submatrices, module the couplings $g_{ij}$, are fixed by the previous requirements~(\ref{YIH}) giving mass scales that range from $10^{-1}-10^{3}$ MeV.
Therefore, to obtain the tiny neutrino mass the scale of the lower block diagonal $3\times 3$ submatrix must be of order of the highest scale of the theory, i.e. $M_s$.
Indeed, following the same procedure we used in the quark and lepton sector, we find at different scales:
%
%
\bea
{\rm 1~TeV ~ scale  }  ~~&:&~~~~E_{5}\sim 0.654 \nn\\
{\rm 10^{12}~GeV ~ scale  }  ~~&:&~~~~E_{5}\sim 0.754  \nn\\
{\rm \L_{GUT} ~scale }   ~~&:&~~~~E_{5}\sim 2.5 \times 10^{-7}
\label{neutscal1}\eea
which are in the expected range.

\subsubsection*{Mixing Matrices}

Using the above values for the couplings and vev's, we can proceed and evaluate the Cabbibo - Kobayashi - Maskawa Matrix (CKM). For the above vacuum, the matrix is:
\bea
{\rm CKM(1 TeV)} &=&
\left(
\begin{array}{lll}
0.970 & 0.240 & 0.007 \\
0.240 & 0.970 & 0.013 \\
0.010 & 0.011 & 0.999
\end{array}
\right)\eea
that has to be compared with the experimental data \cite{PDG}:
\bea
{\rm CKM(Data)} &=&
\left(
\begin{array}{rrr}
0.97419 \pm 0.00022 & 0.2257\pm 0.0010      & ~~~~~0.00359 \pm 0.00016\\
0.2256 \pm 0.0010      & ~~~~~0.97334\pm 0.00023 & 0.0415 \pm 0.001 \\
0.00874^{+0.00026}_{- 0.00037}
                                         & 0.0407\pm 0.0010      &
0.999133^{+0.000044}_{- 0.000043}
\end{array}
\right)\nn\\
\label{CKMdata}\eea
Similarly, we evaluate the neutrino mixing matrix:
\bea
{\rm U_{Neutrino~Mixing}} &=&
\left(
\begin{array}{rrr}
-0.42-0.23 i & -0.53+0.38 i & -0.19-0.54 i \\
0.69-0.21 i  & -0.34+0.10 i & -0.55+0.17 i \\
0.20-0.44 i  &  0.65              & -0.16-0.55 i
\end{array}
\right)\label{Nmixing}\eea

The mixing matrices at $10^{12}{\rm GeV}$ and $\L_{GUT}$ are:
\bea
&&{\rm CKM}(10^{12}{\rm GeV}) = \left(
\begin{array}{rrr}
0.974 & 0.221 & 0.020 \\
0.221 & 0.975 & 0.003 \\
0.019 & 0.007 & 0.999
\end{array}
\right)\\
&&{\rm U_{Neutrino~Mixing}}(10^{12}{\rm GeV}) =
\left(
\begin{array}{rrr}
0.56-0.47 i   & 0.05-0.01 i & 0.66+0.06 i \\
-0.47+0.36 i & 0.42-0.25 i & 0.61+0.09 i \\
0.29-0.01 i   & 0.86             & -0.31-0.24 i
\end{array}
\right)~~~~~\eea
at $M_s=10^{12}$ GeV, and
\bea
&&{\rm CKM}(\L_{GUT}) =
\left(
\begin{array}{rrr}
0.971 & 0.235 & 0.017 \\
0.235 & 0.971 & 0.002 \\
0.017 & 0.001 & 0.999
\end{array}
\right)\\
&&{\rm U_{Neutrino~Mixing}}(\L_{GUT}) =
\left(
\begin{array}{rrr}
0.82              & 0.11-0.44 i & 0.20+0.24 i \\
-0.38-0.32 i & 0.56-0.12 i & 0.33+0.54 i \\
0.19+0.14 i & -0.05+0.67 i & 0.69
\end{array}
\right)~~~~~
\eea
at $M_s=\L_{GUT}$.

\subsection{Vacuum 4: $a=c=d=1, ~b=f=h=0, ~e=g=2$.}

As we mentioned before, among the eight possible vacua in table \ref{8vacua} we were able to find solutions of the form (\ref{YIH}) only for the first three models. Here we concentrate on the fourth model
in table \ref{8vacua}, i.e. $a=c=d=1, ~b=f=h=0, ~e=g=2$.

In this case the corresponding quark mass matrices $M_U,~M_D$ have the form given in (\ref{Uquarkmass1}), while the mass matrices for the leptons and neutrinos have a different form:
\bea
M_L &=&V_d~\left(
\begin{array}{lll}
l_1 v_{\f_2} &l_2 ~ &l_3 ~ \\
l_4 ~ &l_5 v_{\f_1}  &l_6 v_{\f_1} \\
l_7 ~ &l_8 v_{\f_1}  &l_9 v_{\f_1} \\
\end{array}\right)\label{Leptonmass2}
\eea
where $l_i$ are dimensionless couplings assumed to be of the same order and:
\bea
M_N &\sim& \left(
\begin{array}{llllll}
0 & 0 & 0 & g_{11} V_u E_1 & g_{12} V_u E_1 & g_{13} V_u E_1 \\
0 & 0 & 0 & g_{21} V_u E_2 & g_{22} V_u E_2 & g_{23} V_u E_2 \\
0 & 0 & 0 & g_{31} V_u E_2 & g_{32} V_u E_2 & g_{33} V_u E_2 \\
g_{11} V_u E_1 & g_{21} V_u E_2 & g_{31} V_u E_2 & q_{11}M_sE_4 & q_{12}M_sE_4 & q_{13}M_sE_4 \\
g_{12} V_u E_1 & g_{22} V_u E_2 & g_{32} V_u E_2  & q_{21}M_sE_4 & q_{22}M_sE_4 & q_{23}M_sE_4 \\
g_{13} V_u E_1 & g_{23} V_u E_2 & g_{33} V_u E_2  & q_{31}M_sE_4 & q_{32}M_sE_4 & q_{33}M_sE_4
\end{array}\right)\label{Nmatrix2}
\eea
where $E_1,~E_2 $ are the same dimensionless instantonic contributions that also appear in the $U$-quark mass matrix (\ref{Uquarkmass1}).

We can repeat the same procedure and evaluate the values of the vev's and instantons at various scales.
In details we fix the vev's of the two Higges $V_u,~V_d$ to the values of the masses of the heaviest quarks $m_\t, m_b$ at this scale.
This choice implies that the higher mass scale comes from the Yukawa terms.
In order to evaluate the values for the rest of the unknown parameters, we choose at random the norm of the couplings
$|g_i|,~|q_i|,~|l_i|,~|g_{ij}|,~|q_{ij}|$ in a small interval of $[0.1,0.6]$ and we solve the systems equating
the three eigenvalues of each matrix with the masses of the relevant particles.

To summarize we have computed these values at the scales of 1 TeV, $10^{12}$ GeV and at GUT scale. The results are given in the following table:
%
%
\bea
\centering
\begin{tabular}[htb]{|c|c|c|c|c|c|c|c|c|}
\hline $ ~~M_s ~~$ & $ ~~V_u ~~$  & $ ~~V_d ~~$ & $ ~~v_{\f_1} ~~$ & $ ~~v_{\f_2} ~~$& $ ~~{E_1} ~~$& $ ~~{E_2} ~~$& $ ~~{E_3} ~~$& $ ~~{E_4} ~~$\\
\hline
\hline $1$ TeV & 644000& 8920&   0.62 &   0.34 &  $1.66\times 10^{-6}$& 0.0008& 0.003& 0.35\\
\hline $10^{12}$ GeV &    452960&    3160&      0.53 &  0.52     & $1.54\times 10^{-6}$& 0.0006& 0.004& $3\times 10^{-9}$ \\
\hline $\L_{GUT}$ &          378800&    2440&   0.56 &  0.55   & $1.32\times 10^{-6}$& 0.0006& 0.004& $5\times 10^{-14}$\\
\hline
\end{tabular}\nn
\eea

Notice that we have three scales:  one related to the Yukawa terms $V_{u},~V_{d}$,
 one related to the higher order terms $v_{\f_1},~v_{\f_2}$ and one related to the instanton terms $E_{1},~E_{2},~E_{3}$.

The $E_4$ instanton is much higher than the other instanton contributions
because it appears in the Majorana part (lower right $3\times 3$ submatrix)
of the seesaw neutrino mass matrix~(\ref{Nmatrix2}).

\subsubsection*{Mixing Matrices}


Using the above values for the couplings and vev's, we can proceed and evaluate the Cabbibo - Kobayashi - Maskawa Matrix (CKM). For the above vacuum, the matrix is:
\bea
{\rm CKM(1 TeV)} &=&
\left(
\begin{array}{lll}
0.97323 & 0.22979 & 0.00300 \\
0.22971 & 0.97235 & 0.04200 \\
0.00673 & 0.04157 & 0.99911
\end{array}
\right)\eea
that is in agreement with data (\ref{CKMdata}).
Similarly, we evaluate the neutrino mixing matrix:
\bea
{\rm U_{Neutrino~Mixing}} &=&
\left(
\begin{array}{rrr}
0.484+0.118 i & 0.166-0.687 i & -0.486-0.117 i \\
0.294+0.643 i & 0.001 & 0.295 + 0.642 i \\
-0.5 i & 0.707 & 0.5 i
\end{array}
\right)\label{Nmi1xing}
\eea

For the rest of the scales, the mixing matrices are:
\bea
&&{\rm CKM}(10^{12}{\rm GeV}) = \left(
\begin{array}{rrr}
0.992 & 0.111 & 0.038 \\
0.114 & 0.835 & 0.538 \\
0.028 & 0.538 & 0.842
\end{array}
\right)\\
&&{\rm U_{Neutrino~Mixing}}(10^{12}{\rm GeV}) =
\left(
\begin{array}{rrr}
0.995~ & 0.04-0.04 i & 0.05+0.05 i \\
-0.076 i~ & 0.74+0.24 i & -0.61+0.06 i \\
-0.054~ & 0.56+0.25 i &   0.78
\end{array}
\right)~~~~~\\
\nn\eea
at $M_s=10^{12}$ GeV, and
\bea
&&{\rm CKM}(\L_{GUT}) =
\left(
\begin{array}{rrr}
0.973 & 0.228 & 0.003 \\
0.228 & 0.972 & 0.042 \\
0.006 & 0.041 & 0.999
\end{array}
\right)\\
&&{\rm U_{Neutrino~Mixing}}(\L_{GUT}) =
\left(
\begin{array}{rrr}
-0.43-0.11 i ~&  0.76-0.06 i ~& 0.05-0.46 i \\
-0.07-0.34 i ~& -0.18-0.59 i ~& 0.70 \\
0.82              ~&  0.13-0.11 i~ & 0.02-0.54 i
\end{array}
\right)~~~~~
\eea
at $M_s=\L_{GUT}$.

\subsection{The KST vacua}

In this section we consider a different vacuum which was studied in \cite{kst}.
This is a (almost) free-field vacuum with tadpole cancellation. The gauge group is $U(3)\times Sp(2)\times U(1)\times U(1)'$ times
an additional $SU(2)$ coming from the hidden sector of the vacuum. The massless spectrum contains:
\bea
\begin{tabular}[h]{llcllcllc}
$Q_1,~Q_2,~Q_3$& $:~~(~~1, + 1 , ~~0 ,~~ 0)$\\
$U^c_1$& $:~~(-1, ~~0 , -1 , ~~0)$ && $U^c_2~U^c_3$ & $:~~(-1, ~~0 , ~~0 , -1)$\\
$D^c_1$& $:~~(-1, ~~0 , +1 , ~~0)$ && $D^c_2~D^c_3$ & $:~~(-1, ~~0 , ~~0 , +1)$\\
$L^c_1$& $:~~(~~0, +1 ,~~0 , -1)$ && $L^c_2~L^c_3$ & $:~~(~~0, +1 , -1, ~~0)$\\
$E^c_1,~E^c_2, ~E^c_3$& $:~~(~~0,~~ 0 , +1 , +1)$\\
$N^c_1$ & $:~~(~~0, ~~0 , -1, +1)$ && $N^c_2,~N^c_3$ & $:~~(~~0, ~~0 ,~~ 0, ~~0)$
\end{tabular}\label{vacua2}
\eea
Notice that the two right-handed neutrinos $N^c_2,~N^c_3$ come from the hidden sector and in particular, they are described by an antisymmetric and it's conjugate representation of the hidden $SU(2)$ sector.
The two MSSM Higgses are described by
\bea
H_u~&:&~ (0, -1 , +1 , 0)\nn\\
H_d~&:&~ (0,  +1 , -1 , 0) ~.
\eea
In this vacuum, an instanton $E_1$ and its conjugate $E_1^*$ are needed in order to generate the relevant mass terms for the fermions.
The form of this instantons are:
\bea
E_1~&:&~ (0, 0 , -1 , +1)\nn\\
E_1^*~&:&~ (0, 0 , +1 , -1)
\eea
The quark mass matrices for this vacuum are given by:
\bea
M_U &=&V_u~\left(
\begin{array}{lll}
g_1 ~ & g_2 ~ & g_3 E_1^{*} \\
g_4 ~ & g_5 ~ & g_6 E_1^{*} \\
g_7 ~ & g_8 ~ & g_9 E_1^{*} \\
\end{array}\right)\label{UquarkmassKST}\\
M_D &=&V_d~\left(
\begin{array}{lll}
q_1 ~ & q_2 ~ & q_3 E_1 \\
q_4 ~ & q_5 ~ & q_6 E_1 \\
q_7 ~ & q_8 ~ & q_9 E_1 \\
\end{array}\right)
\eea
while the lepton and neutrino mass matrices are given by:
\bea
&M_L =V_d~\left(
\begin{array}{lll}
l_1 ~ & l_2 ~ & l_3  \\
l_4 ~ & l_5 ~ & l_6 \\
l_7  E_1 & l_8  E_1 & l_9 E_1 \\
\end{array}\right)&\label{LeptonmassKST}\\
%
&M_N = \left(
\begin{array}{llllll}
0 & 0 & 0 & g_{11} V_u ~ & g_{12} V_u E_1^{*} & g_{13} V_u E_1^{*} \\
0 & 0 & 0 & g_{21} V_u ~ & g_{22} V_u E_1^{*} & g_{23} V_u E_1^{*} \\
0 & 0 & V_u^2/{M_s} & g_{31} V_u E_1 & g_{32} V_u         & g_{33} V_u  \\
g_{11} V_u  & g_{21} V_u    & g_{31} V_u E_1 & q_{11}M_sE_1^2 & q_{12}M_sE_1 & q_{13}M_sE_1 \\
g_{12} V_u E_1^{*} & g_{22} V_u E_1^{*} & g_{32} V_u  & q_{21}M_sE_1 & q_{22}M_s        & q_{23}M_s  \\
g_{13} V_u E_1^{*} & g_{23} V_u E_1^{*} & g_{33} V_u  & q_{31}M_sE_1 & q_{32}M_s        & q_{33}M_s
\end{array}\right)&\label{NmatrixKST}
\eea
Notice that in this scenario we have only Yukawas and one instanton term that contribute to the mass matrices.

In order to evaluate the instanton, we fix at random the norm of the couplings
$|g_i |$, $|q_i|$, $|l_i|$, $ |g_{ij} |$, $|q_{ij}|$ to be of the same order and we solve the systems equating the three eigenvalues of each matrix with the masses of the relevant particles.
It is worth noting that we are able to reproduce all the fermion masses for a single value of the instanton $E_1$ except for the neutrino masses.
The results are given in the table:
%
%
\bea
\centering
\begin{tabular}[htb]{|c|c|c|c|c|c|c|c|c|c|c|}
\hline $ ~~M_s ~~$ & $ ~~V_u ~~$  & $ ~~V_d ~~$ &  $ ~~{E_1} ~~$\\
\hline
\hline $1$ TeV             &    644000 & 2230 & 2.191 \\
\hline $10^{12}$ GeV &    452960& 3160 & 3.429 \\
\hline $\L_{GUT}$       &    378800& 2440 & 3.245 \\
\hline
\end{tabular}
\nn\eea
%
The corresponding CKM matrices:
\bea
{\rm CKM(1TeV)} &=&
\left(
\begin{array}{lll}
0.727 & 0.444 & 0.522 \\
0.554 & 0.755 & 0.350 \\
0.403 & 0.481 & 0.777
\end{array}
\right)\\
{\rm CKM}(10^{12}{\rm GeV}) &=&
\left(
\begin{array}{lll}
0.825 & 0.533 & 0.184 \\
0.496 & 0.841 & 0.214 \\
0.269 & 0.085 & 0.959
\end{array}
\right)\\
{\rm CKM}(\L_{GUT}) &=&
\left(
\begin{array}{lll}
0.662 & 0.543 & 0.515 \\
0.554 & 0.675 & 0.486 \\
0.503 & 0.498 & 0.705
\end{array}
\right)
\eea
Due to the small number of parameters (two vev's and only one instanton) we did not succeed in satisfying the CKM constraints with couplings in the range 0.1-0.6.
In this case, we didn't include the neutrino mixing matrices since we have not found
any solution that gives approximately correct values for their masses.

\subsection{A vacuum with five stacks}

In vacua with five stacks of branes we have more possible charge assignments for each MSSM
particle. In particular, it is possible to find configurations in which all quark singlets are different.

This possibility allows for vacua where $M_U,~ M_D$ and $M_L$ are of the form (\ref{form5}).

As an example, we consider a vacuum that could be considered as a 5-stack extension of the original
Madrid vacuum with hypercharge $Y={1\over 6}Q_{\brn a}+{1\over 2}Q_{\brn c}+{1\over 2}Q_{\brn d}+{1\over 2}Q_{\brn e}$.
Our main interest is to focus in a case where are quark and lepton singlets are different.
A vacuum that is free of anomalies and satisfies our requirements is:
\bea
\begin{array}{lll}
Q_1~: ~(~~1, + 1, ~~0, ~~0, ~~0) ,&~~ Q_2~:~(~~1, - 1 ,~~ 0 ,~~ 0, ~~0),&~~Q_3~:~(~~1, - 1 ,~~ 0 , ~~0,~~ 0)\\
U^c_1~:~(-1, ~~0 , -1, ~~0 ,~~0) ,&~~ U^c_2~:~(-1,~~ 0 , ~~0 , -1, ~~0) ,&~~ U^c_3~:~(-1,~~ 0 , ~~0, ~~0 , -1)\\
D^c_1~:~(-1, ~~0 , +1, ~~0 ,~~0),&~~ D^c_2~:~(-1, ~~0 ,~~ 0 , +1, ~~0),&~~ D^c_3~:~(-1, ~~0 ,~~ 0, ~~0 , +1)\\
L^c_1~:~(~~0, -1 , -1 , ~~0,~~0) ,&~~ L^c_2~:~(~~0, -1 , ~~0, -1,~~0),&~~ L^c_3~:~(~~0, -1 , ~~0,~~ 0,-1)\\
E^c_1~:~(~~0,~~0, ~~0 , ~~1 , ~~1) ,&~~ E^c_2~:~(~~0,~~0, ~~0 , ~~1, ~~1) ,&~~ E^c_3~:~(~~0, ~~0,~~2 ,~~ 0, ~~0)\\
N^c_1~:~(~~0, ~~2 , ~~0 , ~~0,~~0) ,&~~ N^c_2~:~(~~0, ~~2 ,~~0, ~~0, ~~0),& ~~N^c_3~:~(~~0, ~~2 ,~~0,~~ 0, ~~0)\\
\end{array}\nn\\
\label{vacua3}
\eea
The two MSSM Higgses are described by
\bea
H_u~&:&~ (~~0, -1 , +1 , ~~0,~~0)\nn\\
H_d~&:&~ (~~0, -1 , -1 , ~~0,~~0)
\eea
and we consider two additional scalars with zero hypercharge $\f_1,\f_2$, coming from the non chiral part of the spectrum:
\bea
\f_1~&:&~ (~~0, ~~0 , -1 , +1,~~0)\nn\\
\f_2~&:&~ (~~0,~~ 0 , +1 , -1,~~0) ~.
\eea

The corresponding mass matrices for the quarks and leptons have the form:
\bea
M_U &=&V_u~\left(
\begin{array}{lll}
g_1 ~        &g_2 v_{\f_1} &g_3 E_1 \\
g_4 E_2 &g_5 E_3        &g_6 E_4 \\
g_7 E_2 &g_8 E_3        &g_9 E_4 \\
\end{array}\right)\label{Uquarkmass5stack}\\
M_D &=&V_d~\left(
\begin{array}{lll}
q_1 ~        &q_2 v_{\f_2} &q_3 E_5 \\
q_4 E_2 &q_5 E_6      &q_6 E_7 \\
q_7 E_2 &q_8 E_6       &q_9E_7 \\
\end{array}\right)\\
M_L &=&V_d~\left(
\begin{array}{lll}
l_1 ~      &l_2 v_{\f_2} &l_3 E_8 \\
l_4 E_5 &l_5 E_9      &l_6 E_{10} \\
l_7 E_5 &l_8 E_9      &l_9E_{10} \\
\end{array}\right)\label{Leptonmass5stack}
\eea
and the neutrino mass matrix have the form:
\bea
M_N &=& \left(
\begin{array}{llllll}
0 & 0 & 0 & g_{11} V_u E_2 & g_{12} V_u E_2 & g_{13} V_u E_2 \\
0 & 0 & 0 & g_{21} V_u E_2 & g_{22} V_u E_2 & g_{23} V_u E_2 \\
0 & 0 & 0 & g_{31} V_u E_3 & g_{32} V_u E_3 & g_{33} V_u E_3 \\
g_{11} V_u E_2 & g_{21} V_u E_2 & g_{31} V_u E_3 & q_{11}M_sE_{11} & q_{12}M_sE_{11} & q_{13}M_sE_{11} \\
g_{12} V_u E_2 & g_{22} V_u E_2 & g_{32} V_u E_3  & q_{21}M_sE_{11} & q_{22}M_sE_{11} & q_{23}M_sE_{11} \\
g_{13} V_u E_2 & g_{23} V_u E_2 & g_{33} V_u E_3  & q_{31}M_sE_{11} & q_{32}M_sE_{11} & q_{33}M_sE_{11}
\end{array}\right)\label{Nmatrix5stack}
\eea
where again $g_i,~q_i$ and $l_i$ are dimensionless couplings assumed to be of the same order.

Following the same procedure that was described above we evaluate the higher order and instanton terms at different
scales by equating the eigenvalues of the mass matrices to the values of the running masses computed at that scale.
Finally, we find solutions where the heavy quark masses are coming from the Yukawa terms, the middle quark masses
are coming from the instantonic terms and the light quark masses are given by higher order terms:
\bea
&&E_1\sim E_2\sim E_3\sim E_4 \sim m_c/V_u 
\nn\\
&&E_5\sim E_6\sim E_7\sim E_8\sim E_9\sim E_{10}\sim m_s/V_b 
\nn\\
&&v_{\phi_1} \sim m_u/V_u 
\nn\\
&&v_{\phi_2} \sim m_d/V_b 
\label{YIH5}\eea
where $m_u,~m_d,~m_c,~m_s,~m_t,~m_b$ the masses of the corresponding quarks. Notice that these solutions are valid at all scales.


On the other hand, the remaining instanton $E_{11}$ that appear in the neutrino
mass matrix is fixed to:
\bea
{\rm 1~TeV ~ scale  }  ~~&:&~~~~E_{11}\sim 0.654 \nn\\
{\rm 10^{12}~GeV ~ scale  }  ~~&:&~~~~E_{11}\sim 0.754  \nn\\
{\rm \L_{GUT} ~scale }   ~~&:&~~~~E_{11}\sim 2.5 \times 10^{-7}
\eea

The corresponding mixing matrices are:
\bea
&&{\rm CKM(1TeV)} =
\left(
\begin{array}{lll}
0.972 & 0.241 & 0.007 \\
0.241 & 0.975 & 0.014 \\
0.013 & 0.011 & 0.999
\end{array}
\right)\\
&&{\rm U_{Neutrino~Mixing}(1TeV)} =
\left(
\begin{array}{rrr}
-0.44-0.23 i & -0.54+0.37 i & -0.20-0.52 i \\
0.69-0.19 i  & -0.32+0.11 i & -0.54+0.17 i \\
0.20-0.45 i  &  0.66              & -0.16-0.56 i
\end{array}
\right)~~~~
\eea
at 1TeV,
\bea&&{\rm CKM(10^{12} GeV)} =
\left(
\begin{array}{lll}
0.975 & 0.219 & 0.021 \\
0.221 & 0.975 & 0.003 \\
0.018 & 0.007 & 0.999
\end{array}
\right)\\
&&{\rm U_{Neutrino~Mixing}(10^{12} GeV) }=
\left(
\begin{array}{rrr}
0.56-0.46 i   & 0.05-0.02 i & 0.67+0.06 i \\
-0.48+0.36 i & 0.43-0.25 i & 0.62+0.09 i \\
0.29-0.02 i   & 0.85             & -0.32-0.25 i
\end{array}
\right)~~~~\eea
at $10^{12}$ GeV and
\bea
&&{\rm CKM(\L_{GUT})} =
\left(
\begin{array}{lll}
0.973 & 0.234 & 0.015 \\
0.236 & 0.971 & 0.003 \\
0.017 & 0.001 & 0.999
\end{array}
\right)\\
&&{\rm U_{Neutrino~Mixing}}(\L_{GUT}) =
\left(
\begin{array}{rrr}
0.81              & 0.12-0.44 i & 0.21+0.25 i \\
-0.35-0.32 i & 0.56-0.12 i & 0.32+0.55 i \\
0.17+0.13 i & -0.06+0.66 i & 0.69
\end{array}
\right)~~~~\eea
at $\L_{GUT}$. Notice that in this cases, the CKM matrices are very close to the data.

\section{Branes at singularities and $Z_3$ symmetry}

Singularities of compactification manifolds may carry discrete symmetries that can be considered are gauged.
The reason is that such symmetries are remnants of gauge symmetry broken by gauge fluxes trapped in the collapsing cycles.
Such symmetries have important consequences for Yukawa couplings.

We will indicate this in a $Z_3$ example\footnote{Several quasi-realistic D-brane configurations at
$Z_3$ singularities were first analyzed in \cite{bu2}.} where the $Z_3$ symmetry acts on the doublet-triplets but not on the antiquarks
that correspond to strings ending on other branes\footnote{Related discrete symmetries like $S_3$ have been used in \cite{family} in order
to determine masses and mixings in the SM. The difference here is that the action of the symmetry is not left-right  symmetric.
Similarly, a $Z_3$ grading of mass matrices in F-theory compactifications was discussed very recently in \cite{vafa}. }.

In the presence of a $Z_3$, symmetry that mass matrix of Up and Down quarks is of the form (\ref{form4}).
Such a mass matrix has two zero eigenvalues. To give masses to the massless quarks the $Z_3$ must be broken. This can
happen by moving slightly the moduli that control the collapsed cycle. In the case of the standard $Z_3$ orbifold singularity, these are the
27 twisted moduli.
To classify deformations away from $Z_3$ invariance we introduce the generating $Z_3$ transformation $g$, with $g^3=1$
as an action on three objects
\be
g\left(\begin{array}{rrr} C_1 \\ C_2\\C_3\end{array}\right)
=
\left(\begin{array}{rrr} 0 & 1 &0\\
0 & 0 &1\\
1 & 0 &0\end{array}\right)\left(\begin{array}{rrr} C_1 \\ C_2\\C_3\end{array}\right)=\left(\begin{array}{rrr} C_2\\ C_3\\C_1\end{array}\right)
\ee
The (unormalized) eigenvectors of this action are
\be
\left(\begin{array}{rrr} 1 \\ 1\\1\end{array}\right)_{\l=1}\sp \left(\begin{array}{rrr} 1
\\ \rho\\\rho^2\end{array}\right)_{\l=\rho}
\sp \left(\begin{array}{rrr} 1 \\ \rho^2\\\rho\end{array}\right)_{\l=\rho^2}
\ee
where $\rho=e^{2\pi i\over 3}$ and the subscripts indicate the eigenvalues.
Out of the two non-invariant eigenvectors we can build a vector that is complex conjugation
invariant, in the sense that it is invariant under the transformation generated
by
$$e=\left(\begin{array}{rrr} 1 & 0 &0\\
0 & 0 &1\\
0 & 1 &0\end{array}\right)$$
We therefore choose a new orthonormal basis for the  two $Z_3$ breaking eigenvectors in terms of
\be
v_0={1\over \sqrt{3}}\left(\begin{array}{rrr} 1 \\ 1\\1\end{array}\right)\sp v_+=
{1\over \sqrt{6}}\left(\begin{array}{rrr} 2 \\ -1\\-1\end{array}\right)\sp v_-= {1\over \sqrt{2}}\left(\begin{array}{rrr} 0 \\ 1\\-1\end{array}\right)
\ee
$v_+$ has eigenvalue $+1$ under the action of $e$ while $v_-$ has eigenvalue $-1$.
We may now parameterize a general mass matrix as
\be
\sum_{ij} ~A_{ij}~v_i\otimes v_j\sp i,j=0,\pm
\ee
From now-on we will assume that the mass matrices are written in the $Z_3$ basis introduced above.

A mass matrix invariant under the $Z_3$ symmetry acting on the doublet-triplets has
$A_{+i}=A_{-i}=0$.

This mass matrix is of type (\ref{form4}) and has two zero mass eigenvalues.
A matrix that breaks the $Z_3$ symmetry but is invariant under $e$ is given by
non zero $A_{+i}$ matrix elements.
Finally  the matrix breaking $Z_3$ and $e$ has non-zero $A_{-i}$ matrix elements.
By tuning moduli appropriately we can arrange the mass matrix to have a hierarchical breaking of the $Z_3$ and $e$ symmetries
\be
M_{ij}=\e^{i-1}A_{ij}
 \ee
where $\e\ll 1$ and all $A_{ij}\sim {\cal O}(1)$.
The small parameter $\e$ controls the breaking of the $Z_3$ and $e$
symmetries\footnote{There is no a priori reason that the parameters
breaking $Z_3$ and $e$ are simply related. More generally we may write
$M_{0i}\sim O(1)$, $M_{+i}\sim {\cal O}(\epsilon_3)$, $M_{-i}\sim {\cal O}(\epsilon_e)$
where $\e_3<<1$ is controlling $Z_3$ symmetry breaking and $\e_e<<1$ the $e$-symmetry breaking.
In this case the matrix B has the form
 $$B\sim \left(\begin{array}{ccc} {\cal O}(1) & {\cal O}(\e_3) &{\cal O}(\e_e)\\
{\cal O}(\e_3) & {\cal O}(\e_3^2) &{\cal O}(\e_3\e_e)\\
{\cal O}(\e_e) & {\cal O}(\e_3\e_e) &{\cal O}(\e_e^2)\end{array}\right)$$.}.
In the sequel we take the parameters $A_{ij}$ to be real for simplicity. Therefore we will give up
on explaining the size of the  the CP violation parameters of the SM.
It was claimed recently that under a natural measure in the space of KM matrices  the SM CP violation is generic, \cite{gp}.
If this is correct, then obtaining the CP violation of the SM does not need any further fine-tuning.

Setting $B_{ij}=(AA^T)_{ij}=B_{ji}$ then
\be
MM^{\dagger}=B_{ij}~\e^{(i+j-2)}
\ee
The eigenvalues  for $\e\ll 1$ are
\be
m_0^2=B_{00}+{\cal O}(\e^2)\sp m_1^2={(B_{00}B_{++}-B_{0+}^2)^2\over B_{00}}\e^2+{\cal O}(\e^4)
\ee
\be
m_2^2={(det~B)\over (B_{00}B_{++}-B_{0+}^2)^2}\e^4+{\cal O}(\e^6)
\ee
We therefore generate a natural hierarchy of the masses, $\epsilon_u\sim \l^4$, $\epsilon_d\sim \l^2$, with $\l\sim 0.8$.

In particular to generate the proper hierarchy of masses for the up-type quarks $\e_u=\l^4$
while for the down-type quarks $\e_d=\l^2$ with $\l\simeq 0.22$, \cite{rev-m}.

\subsection{The CKM mixing matrix}

To calculate the mixing matrix we need not only the eigenvalues but also the eigenvectors we present them below
First, the eigenvalues are

\be
m_0^2=B_{00}+\e^2{B_{0+}^2\over B_{00}}+\e^4{B_{0+}^2(B_{00}B_{++}-B_{0+}^2)+B_{00}^2B_{0-}^2\over B_{00}^3}
+{\cal O}(\e^6)
\ee
\be
m_{1}^2={(B_{00}B_{++}-B_{0+}^2)\over B_{00}}\e^2+{B_{0+}^2(B_{00}B_{++}-B_{0+}^2)^2-B_{00}^2(B_{00}B_{+-}-B_{0+}B_{0-})^2
\over B_{00}^3(B_{00}B_{++}-B_{0+}^2)}\e^4+{\cal O}(\e^6)
\ee
\be
{m_2^2\over det(B)}={\e^4\over (B_{00}B_{++}-B_{0+}^2)}+{(B_{0+}B_{0-}-B_{00}B_{+-})^2\over (B_{00}B_{++}-B_{0+}^2)^3}\e^6
+{\cal O}(\e^{8})
\ee

The normalized eigenvector $\xi^0$ for the $m_0$ eigenvalue is
\be
\xi^0=\left(1-{\e^2\over 2}{B_{0+}^2\over B_{00}^2}, ~\e{B_{0+}\over B_{00}},~ \e^2{B_{0-}\over B_{00}}\right)+\cdots
\ee
while for the $m_1$ eigenvalue it is
\be
\xi^1=\left(-\e{B_{0+}\over B_{00}},1-{\e^2\over 2}{B_{0+}^2(B_{00}B_{++}-B_{0+}^2)^2+B_{00}^2(B_{00}B_{+-}-B_{0+}B_{0-})^2
\over B_{00}^2(B_{00}B_{++}-B_{0+}^2)^2},\right.
\ee
$$
\left.  \e{B_{00}B_{+-}-B_{0+}B_{0-} \over B_{00}B_{++}-B_{0+}^2}\right)+\cdots
$$
Finally for the $m_2$ eigenvalue we obtain
\be
\xi^2=\left(-\e^2{B_{0-}B_{++}-B_{0+}B_{+-} \over B_{00}B_{++}-B_{0+}^2},-\e{B_{00}B_{+-}-B_{0+}B_{0-} \over B_{00}B_{++}-B_{0+}^2} ,1-{\e^2\over 2}{(B_{00}B_{+-}-B_{0+}B_{0-})^2 \over (B_{00}B_{++}-B_{0+}^2)^2}\right)+\cdots
\ee
These eigenvectors are orthonormal to order $O(\e^2)$.

The associated unitary matrix that diagonalizes the mass matrix can be parameterized in terms of three parameters ($a,b,c$) and $\e$ to order $O(\e^2)$
\be
U=\left(\matrix{ 1-{a^2\over 2}\e^2 & a\e & b\e^2 \cr
-a\e &1-{a^2+c^2\over 2}\e^2 & c\e  \cr
(ac-b)\e^2& -c\e &1-{c^2\over 2}\e^2
}\right)
\ee
both for the Up and the Down quarks.

We may now evaluate the CKM matrix to be:
\bea
V_{CKM}=U^\dagger_U U_D&=&
\left(
\begin{array}{ccc}
1+ a_d a_u \epsilon _d \epsilon _u & a_d \epsilon _d-a_u \epsilon _u & -a_u c_d \epsilon _d \epsilon _u \\
a_u \epsilon _u-a_d \epsilon _d &1+ \left(a_d a_u+c_d c_u\right) \epsilon _d \epsilon _u & c_d \epsilon _d-c_u \epsilon _u \\
-a_d c_u \epsilon _d \epsilon _u & c_u \epsilon _u-c_d \epsilon _d &1+ c_d c_u \epsilon _d \epsilon _u
\end{array}
\right)\nn\\
&=&\left(
\begin{array}{ccc}
1-\frac{1}{2} \lambda ^4 a_d^2 & \lambda ^2 a_d-\lambda ^4 a_u & \lambda ^4 b_d \\
\lambda ^4 a_u-\lambda ^2 a_d & 1-\frac{1}{2} \lambda ^4 \left(a_d^2+c_d^2\right) & \lambda ^2 c_d-\lambda ^4 c_u \\
\lambda ^4 \left(a_d c_d-b_d\right) & \lambda ^4 c_u-\lambda ^2 c_d & 1-\frac{1}{2} \lambda ^4 c_d^2
\end{array}
\right)
\eea
where $\e_u=\l^4,~\e_d=\l^2,~\l\sim 0.22$. If now we assume $a_u<<1$, $c_d<<1$ and in addition:
$a_d\sim 5, ~b_d\sim 1,~c_u\sim 10$, the CKM becomes:
\bea
V_{CKM}&=&\left(
\begin{array}{ccc}
1-\frac{1}{2} \lambda ^4 a_d^2 & \lambda ^2 a_d & \lambda ^4 b_d \\
-\lambda ^2 a_d & 1-\frac{1}{2} \lambda ^4 a_d^2 & -\lambda ^4 c_u \\
\lambda ^4 \left(a_d c_d-b_d\right) & \lambda ^4 c_u & 1
\end{array}
\right)
=\left(
\begin{array}{ccc}
0.970 & 0.242 & 0.0023 \\
-0.242 & 0.970 & -0.023 \\
-0.0023 & 0.023 & 1
\end{array}
\right)~~~~~~
\eea
This is in absolute value close to what is  measured in experiments.

\section{Correlations with experimentally unfavorable couplings.}

There are several renormalizable superpotential couplings that are allowed by the MSSM gauge symmetries but which are severely constrained by data.
They include couplings that violate lepton and baryon number as well as couplings that are otherwise acceptable but may create problems with the 
hierarchy of masses like the $\mu$-term.
Such couplings are listed below 
\be
\n^c ~,~~ H_uH_d ~,~~ \n^c \n^c \n^c ~,~~ H_dH_u \n^c ~,~~
\label{bb}
\ee
\be
d^c d^c u^c ~,~~ Qu^cL ~,~~ Qd^c L ~,~~  LLl^c ~,~~ H_dH_d l^c
\label{bad}\ee
where we have omitted indexes related to the different families.

The first term in (\ref{bb}) generates a tadpole for the $\nu^c$ indicating that the right-handed sneutrino is non-trivial in the vacuum.
This is not necessary problematic, although it does lead to a reanalysis of the higgs potential and the allowed minima (see for example 
\cite{kst}).
  
The second term is the well-known $\mu$-term. Its only problem is that for models with a large characteristic scale,  
its natural size is the same scale and therefore the EW Higgs doublets are heavy, unless the theory is fine-tuned. 
Its unconstrained presence is a problem only for vacua with a string scale of the order of the GUT scale or an intermediate scale.

The third term in (\ref{bb}) is not necessarily problematic, but in the case where  the right-handed sneutrino  has a vev
it affects the Higgs potential minimization and needs to be taken into account.
The fourth term vanishes identically if we have only a pair of Higgs doublets.
The reason is the antisymmetry of the relevant SU(2) invariants and the symmetry of the superpotential couplings:
$\e_{ab}H_d^aH_d^b l^c$.

In (\ref{bad}) all terms are potentially highly problematic.
The first term violates baryon number, the next two violate both baryon and lepton number while the last two violate lepton number.
In the presence of a single $H_d$, the last term vanishes by antisymmetry.
Typically, in phenomenological models  a discrete  R symmetry is invoked to exclude them from the superpotential.

In many orientifold constructions, such terms are excluded due to one 
or more of the several U(1) (typically anomalous) gauge symmetries present.
There are two possibilities in this direction 

(a) The symmetry that forbids them is ``non-anomalous". This means that the associated gauge boson 
does not mix with string theory axions. This condition is more general than the vanishing of four-dimensional 
mixed gauge anomalies, \cite{u1mass,pascal}. Sometimes this is the case with the gauge B-L symmetry.
In such cases this symmetry must be broken spontaneously by the Higgs effect for the model to not be in gross contradiction with data.
Such a symmetry breaking may generate the unwanted terms in (\ref{bb},\ref{bad}) and may render the vacuum experimentally untenable.
 
(b) The symmetry  that forbids them is ``anomalous". This means that the associated gauge boson
mixes with string theory axions. This guarantees that the associated global symmetry, 
typically unbroken in perturbation theory is violated 
by instanton effects. These may be due to standard gauge theory instantons or stringy instantons.
Instanton effects may leave a discrete part of the symmetry unbroken, and this will may play the role of the R-symmetry.

We have nothing more to say about the case (a), but we do for case (b).
The reason is that we have assumed already in the models we analyze, that some instanton effects do appear in order to provide contributions 
to specific Yukawa couplings.
If an unwanted coupling in the superpotential has the same violation of U(1) charges as a term that has a non-zero instanton contribution, then 
there is a non-trivial instanton contribution for this term. Moreover the strengths of such contributions are related as both contributions differ 
only from the disc correlator that contributes (see \cite{su5}). 

Therefore if one of the terms in (\ref{bb},\ref{bad}) has the same charge violation 
as some Yukawa coupling, then this term is generated with a similar strength.
If it does not, then we cannot say for sure if it is generated. 
It may or it may not, and this can be ascertained if we know 
the global structure of the vacuum.

We have analyzed the charge structure of the terms in (\ref{bb},\ref{bad}) in the  eight models in table \ref{8vacua}.
We have found that of all the terms in (\ref{bb},\ref{bad}), only
the $\mu$-term shares the same charge structure as specific Yukawa couplings, and this is true for all all 8 models. 
The relevant Yukawa couplings are  $Q_{2,3}u^c_1H_u$ and $Q_{2,3}d^c_1H_d$ and they share the same instanton $E_1$ with the $\mu$ term.

As we have  found in the hierarchical solutions (\ref{YIH}, \ref{YIH5}), that 
\bea
E_1 \sim m_c/m_t
\eea
we conclude that in such vacua the $\mu$ term is present but its size is
 suppressed by two or three order of magnitudes compared to the characteristic scale of the vacuum.
Therefore for vacua with a string scale $M_s\sim 1-100$ TeV the $\mu$ term can have a natural size. 
For a higher string scale an independent symmetry is needed in order to suppress the size of the $\mu$ term. 

On the other hand,  none of the bad terms in (\ref{bad}) is necessarily generated.

\section{Conclusions}

In this work we have investigated the possibility of generating the hierarchy of the SM Masses using several characteristics features
on orientifold vacua.
In such vacua many of the techniques and ideas used so far in SM building are not always applicable.
This is due to the fact that the charges carried by the SM fields are constrained to satisfy the standard criteria of opens strings.
For example, the doublet triplets are not allowed to carry other gauge charges.
The features we use to generate the mass hierarchies  include
\begin{itemize}
\item The existence of several (anomalous) U(1) symmetries well beyond those present in the SM. Such symmetries are generically present,
and in general provide serious constraints on low-energy couplings

\item The existence of scalars beyond those of the SM that can generate higher-dimension operators
that upon symmetry breaking generate masses suppressed by the string scale.

\item The existence of instanton effects well beyond standard gauge instantons, that can provide small values to couplings
otherwise forbidden by anomalous U(1)s.

\item The possibility to use discrete symmetries that exist at special points in moduli space and which can be broken infinitesimally.

\end{itemize}

With a view of the possibilities we have analyzed bottom-up SM brane configurations with charges that allow the implementation of such mechanisms.

We have classified such configurations and analyzed the promising ones. Our analysis was exploratory and did not analyze concrete orientifold
vacua. We have however shown constructively that the SM mass matrices and mixings, can be accommodated in several configurations
with couplings of O(1).
The outcome of this exercise is a list of brane configurations that seem promising for generating the SM mass hierarchy.

A direct next step is to search for such configurations in the master list of top-down models produced in \cite{adks}. This is under way.

\vspace{.7 in}
\addcontentsline{toc}{section}{Acknowledgments}

\noindent {\bf Acknowledgements} \newline

We are grateful to F. Fucito, G. Leontaris, S. Raby, B. Schellekens and N. Vlachos for
valuable discussions.

This work was  partially supported by  a European Union grant FP7-REGPOT-2008-1-CreteHEPCosmo-228644,
 an ANR grant NT05-1-41861 and a CNRS PICS grant \# 4172.

Elias Kiritsis is on leave of absence from APC, Universit\'e Paris 7, (UMR du CNRS 7164).

\vskip 2cm

\addcontentsline{toc}{section}{Note added}
\noindent {\Large\bf Note Added} \newline

During the completion of this work we became aware of reference \cite{leontaris}
where a similar idea was pursued in a somewhat more restricted context.
After the completion of this work, reference \cite{Cvetic:2009yh} appeared where similar ideas were explored. 
It partly overlaps with the present work.

\newpage

\appendix
\renewcommand{\theequation}{\thesection.\arabic{equation}}
\addcontentsline{toc}{section}{Appendices}
\section*{APPENDIX}
\bigskip\appendix


\section{Masses at various scales}\label{massesData}

In this section, we provide the masses of the SM particles in a supersymmetric framework, for various scales and for $\tan\beta \sim 50$ \cite{Xing:2007fb}.

\begin{center}
\begin{tabular}{rlrrrr}
\hline $particles$&& $~~~~~~\m=m_Z$
&   $~~~~~~\m=1$ TeV&   $~~~~~~\m=10^{12}$ GeV&      $~~~~~~\m=2\times 10^{16}$ GeV    \\
\hline \hline
u&&$1.27\pm0.42$ & $1.15\pm0.38$& $0.62\pm0.22$ &   $0.48\pm0.17$\\
d&&$ 2.90\pm1.19 $ &$  2.20\pm0.90 $& $0.69\pm0.17$  & $ 0.51\pm0.22 $ \\
c&&$ 619 \pm 84 $ &$  557 \pm 77 $& $304 \pm 45$ & $ 237 \pm 37 $ \\
s&&$  55 \pm 15 $ &$  42 \pm 12 $& $13\pm 4$  & $ 10 \pm 3 $ \\
t&&$ 171700 \pm 13000 $ &$  161000 \pm 3700 $& $ 113200 \pm 77000  $  & $ 94700 \pm 80000  $ \\
b&&$ 2890 \pm 110 $ &$  2230 \pm 80 $& $790\pm 50$  & $ 610 \pm 40 $ \\
e&&$ 0.486 $ &$  0.418 $& $ 0.235 $  & $  0.206 $ \\
$\m$&&$ 102.751 $ &$  88.331 $& $49.75 $ & $  43.50 $ \\
$\t$&&$ 1746.24 $ &$  1502.25 $& $875.31$ & $  773.44 $ \\
$\n_e$ && $10^{-9}$ & $ 10^{-9}    $ & $10^{-9}$ & $  10^{-9}   $ \\
$\n_\m$ && $9 \times10^{-9} $ & $  9.38 \times10^{-9}    $ &  $  1.0 \times10^{-8} $ & $  1.0 \times10^{-8}  $ \\
$\n_\t$ && $5.08\times 10^{-8}$ & $ 5.28\times 10^{-8}    $ & $  5.74 \times10^{-8} $ & $   5.74\times 10^{-8} $ \\
\hline
\end{tabular}
\end{center}

\section{D-brane embeddings}

\subsection{Three stacks: the $U(3)\times U(2)\times
U(1)$ vacua}\label{3branes}

There are two possible ways to embed the SM in this D-brane
system of three stacks, \cite{AD}:
\bea
\begin{array}{lll}
\begin{array}{lll}\nonumber
&Y=&-{1\over 3}Q_{\brn a}-{1\over 2}Q_{\brn
b}\\ \\
&Q:&~~~(~~1,~~1,~~0)\\\nonumber
&u^c:&~~~(~~2,~~0,~~0)\\ \nonumber
&d^c:&~~~(-1,~~0,\pm 1)\\\nonumber
&L:&~~~(~~0,-1,\pm 1)\\ \nonumber
&l^c:&~~~(~~0,~~2,~~0)\\ \nonumber
&H:&~~~(~~0,~~1,\pm 1)\\ \nonumber
&H':&~~~(~~0,-1,\pm 1)
\end{array}
&~~~~~~~~~~~~&
\begin{array}{lll}\nonumber
&Y=&~~~{1\over 6}Q_{\brn a}+{1\over 2}Q_{\brn c} \\ \\
&Q:&~~~(~~1,\pm 1,~~0)\\\nonumber
&u^c:&~~~(-1,~~0,-1)\\ \nonumber
&d^c:&~~~(~~2,~~0,~~0)~~~{\rm or}~~~(-1,~~0,~~1)\\ \nonumber
&L:&~~~(~~0,\pm 1,-1)\\ \nonumber
&l^c:&~~~(~~0,~~0,~~2)\\\nonumber
&H:&~~~(~~0,\pm 1,~~1)\\ \nonumber
&H':&~~~(~~0,\pm 1,-1)
\end{array}
\end{array}
\eea
The three numbers in each parenthesis $(q_3, q_2, q_1)$ denote the corresponding $U(1)$ charges
of each particle. The $\pm$ sign is related to the freedom to choose the charge under the $U(1)_2$ since the
corresponding gauge boson does not contribute to the hypercharge. The $2$ denotes antisymmetric/ symmetric
representations for the non-abelian/abelian factors respectively.

\subsection{Four stacks: $U(3)\times U(2)\times U(1)\times U(1)'$ vacua}\label{4branes}

In this section, we study four-stack realizations of the SM.
We continue with the statistics of fours-stack vacua \cite{adks}.

\subsubsection*{Hypercharge $Y=-{1\over 3}Q_{\brn a}-{1\over 2}Q_{\brn b}
+Q_{\brn d}$}\label{Xis0}

The corresponding charge assignments are:
\bea
&Q:&~~~(~~1,-1,~~0,~~0)\nn\\
&U^c:&~~~(~~2,~~0,~~0,~~0)~~~{\rm or}~~~(-1,~~0,~~0,~~1)\nn\\
&D^c:&~~~(-1,~~0,\pm 1,~~0)\nn\\
&L:&~~~(~~0,~~1,\pm 1,~~0)~~~{\rm or}~~~(~~0,-1,~~0,-1)\nn\\
&E^c:&~~~(~~0,-2,~~0,~~0)~~~{\rm or}~~~(~~0,~~0,\pm 1,-1)\nn\\
&N^c:&~~~(~~0,~~0,\pm 2,~~0)\nn\\
&H_u:&~~~(~~0,-1,\pm 1,~~0)~~~{\rm or}~~~(~~0,~~1,~~0,-1)\nn\\
&H_d:&~~~(~~0,-1,~~0,~~1)~~~{\rm or}~~~(~~0,~~1,\pm 1,~~0) \nn\eea

\subsubsection*{Hypercharge
$Y={2\over 3}Q_{\brn a}+{1\over 2}Q_{\brn b}+Q_{\brn c}$}\label{Xis1}

The corresponding charge assignments are:
\bea
&Q:&~~~(~~1,-1,~~0,~~0)\nn\\
&U^c:&~~~(-1,~~0,~~0,\pm 1)\nn\\
&D^c:&~~~(-1,~~0,~~1,~~0)\nn\\
&L:&~~~(~~0,-1,~~0,\pm 1)~~~{\rm or}~~~(~~0,~~1,-1,~~0)\nn\\
&E^c:&~~~(~~0,~~2,~~0,~~0)~~~{\rm or}~~~(~~0,~~0,~~1,\pm 1)\nn\\
&N^c:&~~~(~~0,~~0,\pm 2,~~0)\nn\\
&H_u:&~~~(~~0,-1,~~1,~~0)~~~{\rm or}~~~(~~0,~~1,~~0,\pm 1)\nn\\
&H_d:&~~~(~~0,-1,~~0,\pm 1)~~~{\rm or}~~~(~~0,~~1,-1,~~0) \nn\eea

\subsubsection*{Hypercharge
$Y={1\over 6}Q_{\brn a}+{1\over 2}Q_{\brn c}-{1\over 2}Q_{\brn d}$}\label{Madrid}

The corresponding charge assignments are:
\bea
&Q:&~~~(~~1,\pm 1,~~0,~~0)\nn\\
&U^c:&~~~(-1,~~0,-1,~~0)~~~{\rm or}~~~(-1,~~0,~~0,~~1)\nn\\
&D^c:&~~~(~~2,~~0,~~0,~~0)~~~{\rm or}~~~(-1,~~0,~~1,~~0)~~~{\rm or}~~~(-1,~~0,~~0,-1)\nn\\
&L:&~~~(~~0,\pm 1,-1,~~0)~~~{\rm or}~~~(~~0,\pm 1,~~0,~~1)\nn\\
&E^c:&~~~(~~0,~~0,~~2,~~0)~~~{\rm or}~~~(~~0,~~0,~~1,-1)~~~{\rm or}~~~(~~0,~~0,~~0,-2)\nn\\
&N^c:&~~~(~~0,\pm 2,~~0,~~0)~~~{\rm or}~~~(~~0,~~0,~~1,~~1)~~~{\rm or}~~~(~~0,~~0,-1,-1)\nn\\
&H_u:&~~~(~~0,\pm 1,~~0,-1)~~~{\rm or}~~~(~~0,\pm 1,~~1,~~0)\nn\\
&H_d:&~~~(~~0,\pm 1,~~0,~~1)~~~{\rm or}~~~(~~0,\pm 1,-1,~~0) \nn\eea

\subsubsection*{Hypercharge
$Y={1\over 6}Q_{\brn a}+{1\over 2}Q_{\brn c}-{3\over 2}Q_{\brn
d}$}\label{Xis1over2plusF}

The corresponding charge assignments are:
\bea
&Q:&~~~(~~1,\pm 1,~~0,~~0)\nn\\
&U^c:&~~~(-1,~~0,-1,~~0)\nn\\
&D^c:&~~~(-1,~~0,~~1,~~0)~~~{\rm or}~~~(~~2,~~0,~~0,~~0)\nn\\
&L:&~~~(~~0,\pm 1,~~1,~~0)\nn\\
&E^c:&~~~(~~0,~~0,-1,~~1)~~~{\rm or}~~~(~~0,~~0,~~2,~~0)\nn\\
&N^c:&~~~(~~0,\pm 2,~~0,~~0)\nn\\
&H_u:&~~~(~~0,\pm 1,~~1,~~0)\nn\\
&H_d:&~~~(~~0,\pm 1,-1,~~0) \nn\eea

\subsubsection*{Hypercharge $Y=-{1\over 3}Q_{\brn a}-{1\over 2}Q_{\brn b}$\label{Xis1over2andCC}}

The corresponding charge assignments are:
\bea
&Q:&~~~(~~1,-1,~~0,~~0)\nn\\
&U^c:&~~~(~~2,~~0,~~0,~~0)\nn\\
&D^c:&~~~(-1,~~0,\pm 1,~~0)~~~{\rm or}~~~(-1,~~0,~~0,\pm 1)\nn\\
&L:&~~~(~~0,-1,\pm 1,~~0)~~~{\rm or}~~~(~~0,-1,~~0,\pm 1)\nn\\
&E^c:&~~~(~~0,-2,~~0,~~0)\nn\\
&N^c:&~~~(~~0,~~0,~~0,\pm 2)~~~(~~0,~~0,\pm 2,~~0)~~~~(~~0,~~0,\pm 1,\pm 1)\nn\\
&H_u:&~~~(~~0,~~1,\pm 1,~~0)\nn\\
&H_d:&~~~(~~0,-1,\pm 1,~~0) \nn\eea

\subsubsection*{Hypercharge
$Y=-{5\over 6}Q_{\brn a}-{Q_{\brn b}}-{1\over 2}Q_{\brn c}+{3\over
2}Q_{\brn d}$ \label{Xisminus1over2}}

The above hypercharge embedding is allowed only in cases where the
right-handed neutrino is coming from the hidden sector. The
corresponding charge assignments are:
\bea
&Q:&~~~(~~1,-1,~~0,~~0)\nn\\
&U^c:&~~~(-1,~~0,~~0,~~1)\nn\\
&D^c:&~~~(-1,~~0,~~1,~~0)\nn\\
&L:&~~~(~~0,-1,~~0,~~1)~~~{\rm or}~~~(~~0,~~1,-1,~~0)\nn\\
&E^c:&~~~(~~0,~~0,-2,~~0)~~~{\rm or}~~~(~~0,~~0,~~1,-1)\nn\\
&H_u:&~~~(~~0,-1,~~1,~~0)~~~{\rm or}~~~(~~0,~~1,~~0,-1)\nn\\
&H_d:&~~~(~~0,-1,~~0,~~1)~~~{\rm or}~~~(~~0,~~1,-1,~~0) \nn\eea

\subsubsection*{Hypercharge
$Y={7\over 6}Q_{\brn a}+{Q_{\brn b}}+{3\over 2}Q_{\brn c}+{1\over
2}Q_{\brn d}$\label{Xis3over2}}

The above hypercharge embedding is allowed only in cases where the
right-handed neutrino is coming from the hidden sector. The
corresponding charge assignments are:
\bea
&Q:&~~~(~~1,-1,~~0,~~0)\nn\\
&U^c:&~~~(-1,~~0,~~0,~~1)\nn\\
&D^c:&~~~(-1,~~0,~~1,~~0)\nn\\
&L:&~~~(~~0,~~1,-1,~~0)~~~{\rm or}~~~(~~0,-1,~~0,~~1)\nn\\
&E^c:&~~~(~~0,~~0,~~0,~~2)~~~{\rm or}~~~(~~0,~~0,~~1,-1)\nn\\
&H_u:&~~~(~~0,-1,~~1,~~0)~~~{\rm or}~~~(~~0,~~1,~~0,-1)\nn\\
&H_d:&~~~(~~0,~~1,-1,~~0)~~~{\rm or}~~~(~~0,-1,~~0,~~1) \nn\eea

\section{Summary of Solutions}

In this section, we present the values of the couplings for two indicative vacua: One with four and one with five stacks of branes:

\subsection*{Four stack vacuum 1: $a=c=d=1, ~b=f=h=0, ~e=g=2$.\label{4bM1}}

As it was presented in the main text, the values of the Yukawa, higher and instantonic terms for scale $\Lambda=1TeV$ are:
\bea
&&v_{\f_1} = 0.62\nn\\
&&v_{\f_2} = 0.34\nn\\
&&E_1  = 1.66 \times10^{-6} \nn\\
&&E_2  = 0.0008 \nn\\
&&E_3  = 0.0038 \nn\\
&&E_4=0.357
\eea
The values of the corresponding couplings are:
\bea
&&\left(
\begin{array}{lll}
g_1 & g_2 & g_3 \\
g_4 & g_5 & g_6 \\
g_7 & g_8 & g_9
\end{array}
\right)
=
\left(
\begin{array}{rrr}
0.25 & 0.25 & -0.25 \\
0.25 & 0.25 & 0.247 \\
0.25 & 0.25 & 0.25
\end{array}
\right)
\nn\\
&&
\left(
\begin{array}{lll}
q_1 & q_2 & q_3 \\
q_4 & q_5 & q_6 \\
q_7 & q_8 & q_9
\end{array}
\right)
=
\left(
\begin{array}{rrr}
0.25 & 0.25 & -0.25 \\
0.41 & -0.43-0.03 i & -0.09+0.49 i \\
0.41 & -0.39-0.02 i &  0.03+0.44 i
\end{array}
\right)
\nn\\
&&
\left(
\begin{array}{lll}
l_1 & ~l_2 & ~l_3 \\
l_4 & ~l_5 & ~l_6 \\
l_7 & ~l_8 & ~l_9
\end{array}
\right)
=
\left(
\begin{array}{rrr}
0.25 & 0.25 & -0.25 \\
-0.25 & 0.25 & 0.25 \\
0.25 & 0.25 & -0.275
\end{array}
\right)
\eea
Similar values for the couplings have been found at higher scales.

\subsection*{Five stack branes\label{5bM1}}

The values for the Yukawa, higher and instantonic terms for all scales are:
\bea
&&V_u  =4 m_t ~,~~~~~~~~ V_d  = 4 m_b ~, \nn\\
&&E_1= E_2= E_3/2= E_4 = m_c/V_u 
\nn\\
&&v_{\phi_1} = 2 m_u/V_u 
\nn\\
&&E_5/2 = E_6/4 = E_7/2 = E_8/2 = E_9 = E_{10} = m_s/V_b  
\nn\\
&&v_{\phi_2} = m_d/V_b  
\eea
and the corresponding couplings:
\bea
&&\left(
\begin{array}{lll}
g_1 & g_2 & g_3 \\
g_4 & g_5 & g_6 \\
g_7 & g_8 & g_9
\end{array}
\right)
=
\left(
\begin{array}{rrr}
0.249 & 0.115 & -0.280 \\
-0.116 & -0.228 & -0.259 \\
-0.141 & -0.119 & -0.135
\end{array}
\right)
\nn\\
&&
\left(
\begin{array}{lll}
q_1 & q_2 & q_3 \\
q_4 & q_5 & q_6 \\
q_7 & q_8 & q_9
\end{array}
\right)
=
\left(
\begin{array}{rrr}
-0.249 & -0.146 & 0.241 \\
0.154  &  0.588 & 0.482 \\
0.675  &  0.114 & 0.128
\end{array}
\right)
\nn\\
&&
\left(
\begin{array}{lll}
l_1 & ~l_2 & ~l_3 \\
l_4 & ~l_5 & ~l_6 \\
l_7 & ~l_8 & ~l_9
\end{array}
\right)
=
\left(
\begin{array}{rrr}
-0.07+0.33 i & 0.23-0.39 i & 0.32+0.05 i \\
0.22+0.02 i & -0.49-0.33 i & 0.50 \\
-0.26+0.05 i & -0.39-0.26 i & 0.16
\end{array}
\right)
\eea
Notice that these values of the couplings give the correct masses at all scales.

\section{Diagonalizing mass matrixes and the Cabbibo - Kobayashi - Maskawa Matrix}
We denote the mass matrices for the quarks as $M_U$ and $M_D$. These are $3\times3$ matrices in the
flavor space and in general they are not hermitian. We can construct a related
hermitian matrix
\bea M_U M_U^\dagger \eea
Being hermitian this matrix is diagonalizable with real eigenvalues and thus it
can be decomposed in the form
\bea M_U M_U^\dagger = U_U D_U^2 U_U^\dagger\eea
where $D^2_U$ is a diagonal matrix with positive eigenvalues and $U_U$ is a unitary
matrix composed of the eigenvectors of $M_U M_U^\dagger$ . We can follow the same procedure
for the down-type Yukawa matrix
\bea M_D M_D^\dagger = U_D D_D^2 U_D^\dagger\eea
where $U_D$ is composed of the eigenvectors of $M_D M_D^\dagger$.
The Cabibbo-Kobayashi-Maskawa (CKM) mixing matrix is
\bea V_{CKM}=U_U^\dagger U_D\eea
The matrix $V_{CKM}$ can have complex elements, but it is possible to remove phases
from $V_{CKM}$ by performing phase rotations of the various quark fields.
\subsection{RGE for the CKM matrix}
The running  CKM matrix elements are obtained by solving the related RGE. The result in the MSSM framework has been computed in~\cite{Das:2000uk}:
\begin{equation}
\left|V_{\alpha\beta}(\mu)\right|=\left\{
\begin{array}{ll}
\left|V_{\alpha\beta}(m_t)\right|\, exp\left[\frac{3}{2}\left( I_t(M_S)+ I_b(M_S)\right)
-\left( \tilde{I}_t(M_S)+ \tilde{I}_b(M_S)\right)\right] & \alpha\beta=ub,cb,tb,ts\\
& \\
\left|V_{\alpha\beta}(m_t)\right| & \textrm{otherwise}
\end{array}
\right.
\end{equation}
where, for each quark $f$, the functions $I_f(M_S)$ is defined as
\begin{equation}
I_f(\mu)=\frac{1}{16\pi^2}\int_{\ln(m_t)}^{\ln(\mu)}y_f^2(t')dt'
\end{equation}
computed for $\mu=M_S$, where $M_S$ is the supersymmetry breaking scale. $y_f$ is the corresponding Yukawa coupling. The function $\tilde{I}_f(M_S)$ is defined as
\begin{equation}
\tilde{I}_f(\mu)=\frac{1}{16\pi^2}\int_{\ln(m_S)}^{\ln(\mu)}y_f^2(t')dt'
\end{equation}
The one loop equations for the two vev's $v_u$ and $v_d$ and for $\tan(\beta)$ get also modified. Anyway from now on we will assume that these quantities are constant functions of the renormalization scale $\mu$.
Under this hypothesis one can estimate the numerical value of the CKM matrix at the unification scale $\mu=M_{GUT}\simeq 10^{16}$ GeV~\cite{Fusaoka:1998vc}:
\begin{equation}
V_{CKM}(M_{GUT})=\left(
\begin{array}{ccc}
0.9754 & 0.2206 & -0.0035i\\
-0.2203i & 0.9745 & 0.0433\\
-0.0032i & -0.0005i & 0.9995
\end{array}
\right)
\end{equation}
\section{Seesaw Comments}
The seesaw mechanism cannot solve the problem in low string scale vacua with $M_s\sim TeV$. In this case, the mass-matrix will look like (to be seen as a $6\times 6$ matrix):
\bea
M_N &\sim& \left(
\begin{array}{ll}
0 & V_u \\
V_u & E_4
\end{array}\right)\label{Nmatrix1}
\eea
with three eigenvalues $\sim E_4$ and three $\sim V_u^2/ E_4$. Notice that here, the instantons are giving the Majorana mass $E_4 \n_R \n_R$ and there is no need of Higgs as in all the other cases.

\section{Mass Matrixes of all eight models of Table 1.}

In this section, we present the mass matrices of the all eight bottom-up models of table \ref{8vacua}.
In all these models the $M_U,~M_D$ mass matrices have the same form:
\bea
M_{U} \sim {V_u} \left (
    \begin {array} {ccc}
                     {1} & v_{\phi_2} & v_{\phi_2} \\
              E_{ {u1}} & E_2 & E_2 \\
          E_{ {u1}} & E_2 & E_2
      \end {array}
     \right)
& ~,~~~~~~~~
M_{D} \sim {V_d} \left (
 \begin {array} {ccc}
                   {1} & v_{\phi_1} & v_{\phi_1} \\
           E_{ {d1}} & E_3 & E_3 \\
       E_{ {d1}} & E_3 & E_3
   \end {array}
  \right)
\eea
They differ only in the leptonic sector and the related matrices are given bellow:
\bea
\begin{array}{ll}
1: ~~~~~~M_{L} \sim {V_d}  \left(
\begin{array}{ccc}
E_4 & v_{\phi_1} & {1} \\
E_4 & v_{\phi_1} & {1} \\
E_4 & v_{\phi_1} & {1}
\end{array}
\right)
& ~,~~~~~~~~ M^{N}_{12} \sim
{V_u}  \left(
\begin{array}{ccc}
E_{1} & E_{1} & E_{1} \\
E_{1} & E_{1} & E_{1} \\
E_{1} & E_{1} & E_{1}
\end{array}
\right) \\
2: ~~~~~~M_{L}  \sim {V_d}  \left(
\begin{array}{ccc}
E_4 & {1} & {1} \\
E_4 & {1} & {1} \\
v_{\phi_1} & v_{\phi_2} & {1}
  v_{\phi_2}
\end{array}
\right)
& ~,~~~~~~~~ M^{N}_{12}  \sim
{V_u}   \left(
\begin{array}{ccc}
E_{1} & E_{1} & E_{1} \\
E_{1} & E_{1} & E_{1} \\
E_2 & E_2 & E_2
\end{array}
\right) \\
3: ~~~~~~M_{L}  \sim {V_d}  \left(
\begin{array}{ccc}
E_4 & v_{\phi_1} & v_{\phi_1} \\
v_{\phi_1} & {1} & {1} \\
v_{\phi_1} & {1} & {1}
\end{array}
\right)
& ~,~~~~~~~~ M^{N}_{12}  \sim
{V_u}   \left(
\begin{array}{ccc}
E_{1} & E_{1} & E_{1} \\
E_2 & E_2 & E_2 \\
E_2 & E_2 & E_2
\end{array}
\right) \\
4: ~~~~~~M_{L}  \sim {V_d}  \left(
\begin{array}{ccc}
v_{\phi_1} & {1} & {1} \\
{1} & v_{\phi_2} & v_{\phi_2} \\
{1} & v_{\phi_2} & v_{\phi_2}
\end{array}
\right)
& ~,~~~~~~~~ M^{N}_{12}  \sim
{V_u}  \left(
\begin{array}{ccc}
E_{1} & E_{1} & E_{1} \\
E_2 & E_2 & E_2 \\
E_2 & E_2 & E_2
\end{array}
\right) \\
5: ~~~~~~M_{L}  \sim {V_d}  \left(
\begin{array}{ccc}
E_4 & E_4 & v_{\phi_1} \\
E_4 & E_4 & v_{\phi_1} \\
v_{\phi_1} & v_{\phi_1} & {1}
\end{array}
\right)
& ~,~~~~~~~~ M^{N}_{12}  \sim
{V_u}  \left(
\begin{array}{ccc}
E_{1} & E_{1} & E_{1} \\
E_{1} & E_{1} & E_{1} \\
E_2 & E_2 & E_2
\end{array}
\right) \\
6: ~~~~~~M_{L}  \sim {V_d}  \left(
\begin{array}{ccc}
v_{\phi_1} & v_{\phi_1} & {1} \\
v_{\phi_1} & v_{\phi_1} & {1} \\
{1} & {1} & v_{\phi_2}
\end{array}
\right)
& ~,~~~~~~~~ M^{N}_{12}  \sim
{V_u}  \left(
\begin{array}{ccc}
E_{1} & E_{1} & E_{1} \\
E_{1} & E_{1} & E_{1} \\
E_2 & E_2 & E_2
\end{array}
\right) \\
7: ~~~~~~M_{L}  \sim {V_d}  \left(
\begin{array}{ccc}
E_4 & E_4 & {1} \\
v_{\phi_1} & v_{\phi_1} & {1}
  v_{\phi_2} \\
v_{\phi_1} & v_{\phi_1} & {1}
  v_{\phi_2}
\end{array}
\right)
& ~,~~~~~~~~ M^{N}_{12}  \sim
{V_u}   \left(
\begin{array}{ccc}
E_{1} & E_{1} & E_{1} \\
E_2 & E_2 & E_2 \\
E_2 & E_2 & E_2
\end{array}
\right) \\
8: ~~~~~~M_{L}  \sim {V_d}  \left(
\begin{array}{ccc}
v_{\phi_1} & {1} & v_{\phi_2} \\
v_{\phi_1} & {1} & v_{\phi_2} \\
v_{\phi_1} & {1} & v_{\phi_2}
\end{array}
\right)
& ~,~~~~~~~~ M^{N}_{12}  \sim
{V_u}   \left(
\begin{array}{ccc}
E_2 & E_2 & E_2 \\
E_2 & E_2 & E_2 \\
E_2 & E_2 & E_2
\end{array}
\right)
\end{array}
\eea
where with $M^N_{12}$ we denote only the upper off-diagonal part of the neutrino mass matrix
since the general form can be written as:
\bea
\left(
\begin{array}{lll}
0&M^N_{12}\\
(M^N_{12})^T & M_s E_5~I_{3\times 3}
\end{array}\right)
\eea
which is the standard form in seesaw mechanism.

\end{document}